\documentclass[journal, 12pt, onecolumn]{IEEE_Transaction}

\usepackage{cite}
\usepackage{comment}
\usepackage{amsmath}
\usepackage{algorithmic}
\usepackage{amssymb}
\usepackage{amsthm}
\usepackage{eqnarray}
\usepackage{graphicx}
\ifCLASSOPTIONonecolumn
\usepackage{setspace}
\fi
\newcommand\blfootnote[1]{%
  \begingroup
  \renewcommand\thefootnote{}\footnote{#1}%
  \addtocounter{footnote}{-1}%
  \endgroup}

\newcommand\numberthis{\addtocounter{equation}{1}\tag{\theequation}}

\newcommand{\subparagraph}{}
\usepackage[compact]{titlesec}
\titlespacing{\section}{0pt}{0.7ex}{0.7ex}
\titlespacing{\subsection}{0pt}{0.7ex}{0.7ex}
\titlespacing{\subsubsection}{0pt}{0.7ex}{0.7ex}

\allowdisplaybreaks

\begin{document}
\title{The Effect of Spatial Correlation on the Performance of Uplink and Downlink Single-Carrier Massive MIMO Systems}
\ifCLASSOPTIONonecolumn
\author{Nader Beigiparast, \IEEEmembership{Student Member,} Gokhan M. Guvensen, \IEEEmembership{Member,}\\ and Ender Ayanoglu, \IEEEmembership{Fellow}%
\thanks{N. Beigiparast and E. Ayanoglu are with the Center for Pervasive Communications and Computing, Dept. EECS, UC Irvine, Irvine, CA, USA, G. M. Guvensen is with the Dept. of EEE, METU, Ankara, Turkey.}}
\else
\author{Nader Beigiparast, \IEEEmembership{Student Member,} Gokhan M. Guvensen, \IEEEmembership{Member,} and Ender Ayanoglu, \IEEEmembership{Fellow}%
\thanks{N. Beigiparast and E. Ayanoglu are with CPCC, Dept. EECS, UC Irvine, CA, USA, G. M. Guvensen is with the Dept. of EEE, METU, Turkey.}}
\fi


\maketitle

\begin{abstract}
\textbf{We present the analysis of a single-carrier massive MIMO system for the frequency selective Gaussian multi-user channel, in both uplink and downlink directions. We develop expressions for the achievable sum rate when there is spatial correlation among antennas at the base station. It is known that the channel matched filter precoder (CMFP) performs the best in a spatially uncorrelated downlink channel. However, we show that, in a spatially correlated downlink channel with two different correlation patterns and at high long-term average power, two other precoders have better performance. For the uplink channel, part of the equivalent noise in the channel goes away, and implementing two conventional equalizers leads to a better performance compared to the channel matched filter equalizer (CMFE). These results are verified for uniform linear and uniform planar arrays. In the latter, due to more correlation, the performance drop with a spatially correlated channel is larger, but the performance gain against channel matched filter precoder or equalizer is also bigger. In highly correlated cases, the performance can be a significant multiple of that of the channel matched filter precoder or equalizers.}

\end{abstract}

\ifCLASSOPTIONonecolumn
\vfill
\fi
\blfootnote{This work was partially supported by NSF under Grant No. 1547155.}
\blfootnote{This paper was partially presented during IEEE VTC2018-Spring.}

\IEEEpeerreviewmaketitle
\section{Introduction}
\label{sec:introduction}

The demands for high-rate wireless communications have been increasing in the past few decades \cite{C1}. Due to this fact, there is a great amount of research on the development of new and efficient schemes to obtain higher rates of information for an increasing number of wireless channel users. Massive multiple-input multiple-output (massive MIMO) is one of the approaches to achieve a higher rate in a wireless system. As a corollary of increasing the number of transmitters at the base station, the transmit power can be designed to be significantly small, since interference decreases at the same rate of the desired signal power, leading to a power-efficient system \cite{C2}. Also, the channel vectors (or matrices) defined between transmitters and receivers become asymptotically orthogonal, allowing multiple users to use the same bandwidth without interfering with each other and achieving a high spectral efficiency in the system \cite{C7}.

In most studies of massive MIMO, the base station is considered to have the perfect knowledge of the channel, since massive MIMO relies on spatial multiplexing which requires good channel knowledge for the base station \cite{C8}. The base station can estimate the channel response to an individual user using the pilots sent by the terminals in the uplink. This process is more challenging in the downlink where the base station can send out pilot waveforms and users' terminals will be able to estimate the channel responses and feed them back to the base station.

In this paper, we assume operation in time division duplex (TDD) mode to take advantage of the reciprocity in the channel, as commonly done in the field.

The work in this paper is focused on single-carrier transmission for massive MIMO applications. This is motivated by a number of recent studies such as \cite{C1}. This work showed that a channel matched filter precoder (CMFP) is optimum for such systems in channels uncorrelated in space. However, unlike \cite{C1}, we wish to determine the performance of such systems for correlated channels, such as in massive MIMO where the presence of correlation in the channel is expected to be high due to space limitations for the antenna elements. Also, we expand our studies to the uplink channel, where equalizers are deployed at the base station side. Using the equalizers in the uplink and precoders in the downlink, we investigate the performance of the massive MIMO channel under the influence of correlation patterns. We would like to emphasize that unlike a conventional channel matched filter, CMFP is placed at the transmitter side \cite{C1}. It is possible to extend this work to orthogonal frequency division multiplexing (OFDM). Although \cite{C1} states single-carrier transmission is better for massive MIMO in spatially uncorrelated channels, we believe this requires further study, particularly in the case of spatially correlated channels.

This paper extends \cite{C1} in the following ways. First, it extends the downlink analysis in \cite{C1} to the uplink. The resulting analysis is different than that for the downlink. Second, it shows that for channels with spatial correlation, CMFP is not necessarily the best precoder in the downlink. This result for spatially correlated channels is contrary to the result in \cite{C1} for spatially uncorrelated channels. Third, it shows that, the uplink analysis eliminates one of the equivalent noise terms present in the downlink. This causes a substantial performance increase for all equalizers under consideration. Fourth, while \cite{C1} considers only CMFP on the downlink it considers five more structures as precoder and equalizers in the downlink and uplink, respectuively. Fifth, it shows that an equalizer for the uplink based on the minimum mean squared error criterion performs the best, in a significant way, among those considered. The investigation in this paper is for downlink and uplink performance of two different spatial correlation models, incorporating different degrees of exponential correlation and taking into account different angles of arrival (AOAs) and both isotropic and nonisotropic arrivals. Finally and sixth, the results are provided for both uniform linear array (ULA) and uniform planar array (UPA) structures.

\section{Downlink Channel Analysis}
\label{sec:downlink}

In the downlink model, a frequency-selective multi-user MIMO (MU-MIMO) channel with $M$ base station antennas and $K$ single-antenna users is considered. We model the channel between the $m$-th transmit antenna and the $k$-th user as a finite impulse response (FIR) filter with $L$ taps. The taps model different delay components. The $l$-th tap can be written as $\sqrt{d_l[k]}h^*_l[m,k]$ where $d_l[k]$ and $h^*_l[m,k]$ correspond to the slow-varying and fast-varying components of the channel respectively, and where $h_l[m,k]$ has a complex Gaussian distribution with zero mean and unit variance ${\cal CN}(0,1)$ \cite{C1}. When the antenna elements or delay components are not correlated, the entries of the matrix consisting of the fast-varying components of the channel, $h^*_l[m,k]$, are independent and identically distributed (i.i.d.). Further, these entries are considered to be fixed while the $T$ symbols are being transmitted. Define $\textbf{y}[i] \triangleq [y_1[i], \cdots, y_K[i]]^T \in \mathbb{C}^K$ and $\textbf{x}[i] \triangleq [x_1[i],\cdots,x_M[i]]^T \in \mathbb{C}^M$ as the vector of received signals at each user and the vector of transmitted signals from each antenna at the base station at time $i=1, 2, \cdots, T$, respectively. Assume that the noise vector $\textbf{n}[i] \triangleq [n_1[i], \cdots, n_K[i]]^T$ is additive white Gaussian noise (AWGN) consisting of i.i.d. components and complex Gaussian distribution. We assume that this distribution has zero mean and unit variance. The variable $s_k[i]$ is the information symbol to be transmitted to the $k$-th user at time $i$. The vector of information symbols is defined as $\textbf{s}[i] \triangleq [s_1[i], \cdots, s_K[i]]^T$. This vector is considered to have i.i.d. $\mathcal{CN}(0,\sqrt{\rho_f})$ components. We also need to define $\textbf{D}_l \triangleq \mathrm{diag}\lbrace d_l[1], \cdots, d_l[K]\rbrace $, and $\textbf{H}_l \in \mathbb{C}^{M\times K}$. The $[m,k]$-th element of ${\bf H}_l$ is $h_l[m,k]$.

In this paper, we will model the channel whose antenna elements have spatial correlation. To this end, we introduce the matrix $\textbf{A} \in \mathbb{C}^{M \times M}$, taking into account all $M$ antenna elements at the base station.The effect of this matrix on the channel will be to modify the channel realization from ${\bf H}_l$ to $\textbf{A}^{1/2} \textbf{H}_l$. A channel whose antenna elements are not spatially correlated will have $\textbf{A} = \textbf{I}_M$. In Section~\ref{sec:simulations}, we will make use of two commonly used spatial correlation models from the literature. Both of these models will have symmetric ${\bf A}$ matrices.

The received signal vector can be written as
\begin{equation}
\textbf{y}[i] = \sum_{l=0}^{L-1} \hat{\textbf{H}}_l^H \textbf{x}[i-l] + \textbf{n}[i],
\label{eqn:1}
\end{equation}
where the channel state information (CSI) matrix is defined as $\hat{\textbf{H}}_l = \textbf{A}^{1/2} \textbf{H}_l \textbf{D}_l^{1/2}$ and the channel power delay profile (PDP) for each user is normalized such that \cite{C1}
\begin{equation}
\sum_{l=0}^{L-1} d_l[k] = 1, \quad k = 1,\ldots, K.
\label{eqn:2}
\end{equation}
Also, $\textbf{x}[i]$ is the transmitted signal vector and it depends on the precoder used in the channel.

\subsection{Channel Matched Filter Precoder (CMFP)}
\label{sec:CMFP}

In wireless communication, the precoding scheme has a significant role.
The CMFP response matrix is the Hermitian of the CSI matrix. Therefore, we specify the transmit symbol vector for the proposed CMFP as \cite{C1}
\begin{equation}
\textbf{x}[i] = \sqrt{\frac{1}{MK}} \sum_{l=0}^{L-1} \hat{\textbf{H}}_l \textbf{s}[i+l].
\label{eqn:3}
\end{equation}
We define the super channel matrix as the multiplication of the precoder matrix and the channel information matrix $\textbf{F}_{(l,l^{\prime})} = \hat{\textbf{H}}_l^H \hat{\textbf{H}}_{l^{\prime}}$. Note that one can obtain $\textbf{F}_{(l,l^{\prime})} [q,{q^\prime}] = \textbf{e}^T_q \textbf{F}_{(l,l^{\prime})} \textbf{e}_{q^\prime}$, where $\textbf{e}_q$ is a vector with all its elements equal to $0$ except the $q$-th one which is $1$.
Using (\ref{eqn:3}) in (\ref{eqn:1}), one can obtain
\begin{equation}
{\bf y}[i] = \frac{1}{MK}\sum_{l=0}^{L-1}\sum_{l'=0}^{L-1}{\bf H}_l^H{\bf H}_{l'}{\bf s}[i+l'-l]+{\bf n}[i]
\end{equation}
and by making the change $b=l-l'$, one can write
\begin{equation}
{\bf y}[i] = \frac{1}{\sqrt{MK}} \sum_{b=1-L}^{L-1} \sum_{\max(0,b)}^{\min(L-1,L-1+b)}{\bf H}_l^H{\bf H}_{l-b}{\bf s}[i-b]+{\bf n}[i] .
\label{eqn:4}
\end{equation}
We define the desired signal as
\begin{equation}
{\bf d}[i] = \frac{1}{\sqrt{MK}}\sum_{l=0}^{L-1}{\mathbb E}[{\bf H}_l^H {\bf H}_l ] {\bf s}[i].
\end{equation}
With this definition, one can write (\ref{eqn:4}) as
\begin{equation}
{\bf y}[i] = {\bf d}[i] + {\bf n}'[i]
\label{eqn:4.5}
\end{equation}
and for user $k$, (\ref{eqn:4.5}) simplifies to
\begin{equation}
y_k[i] = d_k[i] + n_k'[i].
\end{equation}
By an inspection of (\ref{eqn:1}), (\ref{eqn:3}), and (\ref{eqn:4}), the effective noise term can be calculated as
\ifCLASSOPTIONonecolumn
\begin{align*}
n_k^\prime[i] = & \sqrt{\frac{1}{MK}} \sum_{l=0}^{L-1} \bigg(\textbf{F}_{(l,l)}[k,k] - \mathbb{E}\big\{\textbf{F}_{(l,l)}[k,k]\big\} \bigg) s_k[i]
 + \sqrt{\frac{1}{MK}} \sum_{\substack{b=1-L \\ b \neq 0}}^{L-1} \sum_{l=L_1}^{L_2} \textbf{F}_{(l,l-b)}[k,k] s_k[i-b] \\
& + \sqrt{\frac{1}{MK}} \sum_{\substack{q=1 \\ q \neq k}}^{K} \sum_{b=1-L}^{L-1} \sum_{l=L_1}^{L_2} \textbf{F}_{(l,l-b)}[k,q] s_q[i-b]
 + n_k[i].\numberthis
\label{eqn:5}
\end{align*}
\else
\begin{align*}
& n_k^\prime[i] = \sqrt{\frac{1}{MK}} \sum_{l=0}^{L-1} \bigg(\textbf{F}_{(l,l)}[k,k] - \mathbb{E}\big\{\textbf{F}_{(l,l)}[k,k]\big\} \bigg) s_k[i] \\
& + \sqrt{\frac{1}{MK}} \sum_{\substack{b=1-L \\ b \neq 0}}^{L-1} \sum_{l=L_1}^{L_2} \textbf{F}_{(l,l-b)}[k,k] s_k[i-b] \\
& + \sqrt{\frac{1}{MK}} \sum_{\substack{q=1 \\ q \neq k}}^{K} \sum_{b=1-L}^{L-1} \sum_{l=L_1}^{L_2} \textbf{F}_{(l,l-b)}[k,q] s_q[i-b] \\
& + n_k[i].\numberthis
\label{eqn:5}
\end{align*}
\fi
In (\ref{eqn:5}), we defined $L_1 \triangleq \max(b,0)$ and $L_2 \triangleq \min(L-1+b, L-1)$. In (\ref{eqn:5}), the first term can be identified as the additional interference (IF), the second term as the intersymbol interference (ISI), and the third term as multiuser interference (MUI). The final term is AWGN.

Since the effect of the correlation pattern is not present in the power of the desired signal (see Proposition 1), in (\ref{eqn:4}), for the desired signal term, the average power is equal to \cite{C1}
\begin{equation}
S_k = \mathbb{E}_{s_k[i]} \bigg\{\bigg|\sqrt{\frac{1}{MK}} \sum_{l=0}^{L-1} \mathbb{E}\big\{\textbf{F}_{(l,l)}[k,k]\big\} s_k[i]\bigg|^2\bigg\} = \frac{M\rho_f}{K}.
\label{eqn:6}
\end{equation}
The correlation matrix and the average power of the desired signal are independent, meaning that the desired signal power is not affected by the spatial correlation among antennas at the base station. Knowing that all different terms in the effective noise equation are independent from one another (see Proposition 2), the power of $n_k'[i]$, the effective noise in (\ref{eqn:5}), can be calculated as
\begin{equation}
\mathrm{Var}\big(n_k^\prime[i]\big) = \mathrm{tr(\textbf{A}^2)}{M}\rho_f + 1.\label{eqn:7}
\end{equation}
In (\ref{eqn:7}), $\mathrm{tr}(\textbf{A}^2)$ is the trace of the square of the correlation matrix. We will introduce Proposition 1 and Proposition 2 below. These two propositions illustrate how (\ref{eqn:4}) can be used to calculate the average power of the desired signal in (\ref{eqn:6}) and the effective noise power in (\ref{eqn:7}). For a special case in \cite{C1} where the channel does not have spatial correlation, (\ref{eqn:7}) becomes $\rho_f + 1$. Although the average power of the desired signal and the correlation matrix are independent, the effective noise power is affected by the correlation matrix.

We are now interested in characterizing ${\rm tr}({\bf A}^2)$ for the spatially correlated channel. When we consider the correlated case, the entries on the main diagonal of ${\bf A}$ remain the same as the spatially uncorrelated case. On the other hand, other elements have nonzero values when there is correlation and for the models in this work, these values are nonnegative. Therefore, for the spatially correlated channel,
${\rm tr}({\bf A}^2) > M$. For this reason, the effective noise power will increase because of channel correlation. On the other hand, this does not affect the desired signal's average power. The correlation pattern makes the user information rate decrease, but the capacity remains the same. This implies an increase in correlation results in increased difference between the information rate and the capacity.

Each user's information rate can be obtained by $R_k = \frac{1}{2}\log_2 \big(1 + \frac{M^2\rho_f}{K\mathrm{tr}(\textbf{A}^2)\rho_f + KM}\big)$. By considering the fact that each user's information rate is almost equal to the other users', we will have the sum rate as
\begin{equation}
R_{\it sum}(\rho_f, M, K) = \frac{K}{2} \log_2 \bigg(1 + \frac{M\rho_f}{K\frac{\mathrm{tr}(\textbf{A}^2)}{M}\rho_f + K}\bigg).
\label{eqn:8}
\end{equation}

Since the correlation has no effect on the average power of the desired signal, the cooperative sum-capacity is independent of the correlation pattern and it can be written as \cite{C1}
\begin{equation}
C_{\it coop}(\rho_f, M, K) \approx \frac{K}{2}\log_2\big(1 + \frac{M\rho_f}{K}\big).
\label{eqn:9}
\end{equation}
Equations (\ref{eqn:8}) and (\ref{eqn:9}) are the sum rate (information rate) and the upper bound for CMFP, respectively. 
By observing the information rate and the sum-capacity, one can conclude that CMFP may not be a good choice when there is sufficiently significant spatial correlation among the base station antennas. When the channel does not have spatial correlation, the effect of AWGN becomes more pronounced over the effective noise power. As a result, in that case, CMFP appears as a good choice for a precoder. When $\rho_f$ reaches larger values, the information rate begins to saturate until the efective noise power is dominated by the interference terms. As correlation increases, saturation happens earlier and quicker. The effect of any correlation among base station antennas is stronger interference terms that have a bigger effect than AWGN.

\noindent{\bf Proposition 1:} The average power of the signal using CMFP is given as in (\ref{eqn:6}).\\
\noindent{\em Proof:\/}
In order to obtain the average power of the desired signal, we need the following lemma.\\
\noindent{\bf Lemma 1:} Let ${\bf x}$ and ${\bf y}$ be $N$-dimensional zero-mean circularly symmetric complex Gaussian vectors with the autocorrelation and cross-correlation matrices given as ${\bf R}_{\bf x}\triangleq \mathbb{E} \{ {\bf x}{\bf x}^H\}$, ${\bf R}_{\bf y}\triangleq \mathbb{E} \{ {\bf y}{\bf y}^H\}$, and ${\bf R}_{{\bf y}{\bf x}}\triangleq \mathbb{E} \{ {\bf y}{\bf x}^H\}$. Then, the following expectations can be expressed as
\ifCLASSOPTIONonecolumn
\begin{align}
\mathbb{E} \{ {\bf x}^H {\bf A}{\bf y} \} & = {\rm tr}({\bf A}{\bf R}_{{\bf y}{\bf x}})\label{eqn:gmgstar}\\
\mathbb{E}\{|{\bf x}^H{\bf A}{\bf y}|^2\} & =
{\rm tr}({\bf A}^T{\bf R}_{\bf x}{\bf A}{\bf R}_{\bf y}) + {\rm tr}({\bf A}{\bf R}_{\bf yx}){\rm tr}({\bf A}^T{\bf R}_{\bf xy})
\end{align}
\else
\begin{align}
\mathbb{E} \{ {\bf x}^H {\bf A}{\bf y} \} & = {\rm tr}({\bf A}{\bf R}_{{\bf y}{\bf x}})\label{eqn:gmgstar}\\
\mathbb{E}\{|{\bf x}^H{\bf A}{\bf y}|^2\} & = \nonumber\\
& \mkern-18mu {\rm tr}({\bf A}^T{\bf R}_{\bf x}{\bf A}{\bf R}_{\bf y}) + {\rm tr}({\bf A}{\bf R}_{\bf yx}){\rm tr}({\bf A}^T{\bf R}_{\bf xy})
\end{align}
\fi
in which ${\bf A}$ is a real-valued and symmetric $N\times N$ matrix. The proof of this lemma is omitted here due to lack of space.\footnote{One can use the moment generating function to obtain (\ref{eqn:gmgstar}).}

Consider the average power of the desired signal in (\ref{eqn:4}) as
\ifCLASSOPTIONonecolumn
\begin{align}
S_k & = \frac{1}{MK} \mathbb{E}_{s_k[i]} \bigg\{\bigg| \sum_{l=0}^{L-1} \mathbb{E} \big\{\textbf{F}_{(l,l)}[k,k]\big\} s_k[i]\bigg|^2\bigg\} \nonumber\\
& = \frac{1}{MK} \mathbb{E}_{s_k[i]} \bigg\{\sum_{l^\prime=0}^{L-1} s_k^H[i] \mathbb{E} \big\{\textbf{F}_{(l^\prime,l^\prime)}[k,k]\big\}
\sum_{l=0}^{L-1} \mathbb{E} \big\{\textbf{F}_{(l,l)}[k,k]\big\} s_k[i]\bigg\}.
\label{eqn:11}
\end{align}
\else
\begin{align}
S_k & = \frac{1}{MK} \mathbb{E}_{s_k[i]} \bigg\{\bigg| \sum_{l=0}^{L-1} \mathbb{E} \big\{\textbf{F}_{(l,l)}[k,k]\big\} s_k[i]\bigg|^2\bigg\} \nonumber\\
& = \frac{1}{MK} \mathbb{E}_{s_k[i]} \bigg\{\sum_{l^\prime=0}^{L-1} s_k^H[i] \mathbb{E} \big\{\textbf{F}_{(l^\prime,l^\prime)}[k,k]\big\} \nonumber\\
& \qquad\qquad\qquad\qquad\quad\times \sum_{l=0}^{L-1} \mathbb{E} \big\{\textbf{F}_{(l,l)}[k,k]\big\} s_k[i]\bigg\}.
\label{eqn:11}
\end{align}
\fi
We claim that the first multiplier can be written as
\ifCLASSOPTIONonecolumn
\begin{equation}
\sum_{l^\prime=0}^{L-1} s_k^H[i] \mathbb{E} \big\{\textbf{F}_{(l^\prime,l^\prime)} [k,k] \big\} =
\sum_{l^\prime=0}^{L-1} s_k^H[i] \mathbb{E} \big\{\textbf{e}_k^T \textbf{D}_{l^\prime}^{1/2} \textbf{H}_{l^\prime}^H \textbf{A} \textbf{H}_{l^\prime} \textbf{D}_{l^\prime}^{1/2} \textbf{e}_k\big\}.
\label{eqn:12}
\end{equation}
\else
\begin{align}
\sum_{l^\prime=0}^{L-1} s_k^H[i] \mathbb{E} & \big\{\textbf{F}_{(l^\prime,l^\prime)} [k,k] \big\} = \nonumber\\
& \sum_{l^\prime=0}^{L-1} s_k^H[i] \mathbb{E} \big\{\textbf{e}_k^T \textbf{D}_{l^\prime}^{1/2} \textbf{H}_{l^\prime}^H \textbf{A} \textbf{H}_{l^\prime} \textbf{D}_{l^\prime}^{1/2} \textbf{e}_k\big\}.
\label{eqn:12}
\end{align}
\fi
Let's focus on the term $\mathbb{E}\big\{\textbf{F}_{(l,l)}[k,k]\big\}$. If we expand this term, it will be
\begin{equation}
\mathbb{E}\big\{\textbf{F}_{(l,l)}[k,k]\big\} = \mathbb{E}\big\{\textbf{e}_k^T \textbf{D}_l^{1/2} \textbf{H}_l^H \textbf{A} \textbf{H}_l \textbf{D}_l^{1/2} \textbf{e}_k\big\}.
\label{eqn:13}
\end{equation}
Note that using $\textbf{D}_l^{1/2} \textbf{e}_k = \sqrt{d_l[k]} \textbf{e}_k$, we can rewrite the above equation as
\begin{equation}
\mathbb{E}\big\{\textbf{F}_{(l,l)}[k,k]\big\} = \sqrt{d_l[k]} \mathbb{E}\big\{\textbf{h}_l^H [k] \textbf{A} \textbf{h}_l[k] \big\} \sqrt{d_l[k]},
\label{eqn:14}
\end{equation}
where $\textbf{h}_l[k] \triangleq \textbf{H}_l\textbf{e}_k$. Note that $\mathbb{E}\{\textbf{h}_l[k]\} = \mathbb{E}\{\textbf{h}_l^H [k]\} = 0$ and $\mathrm{Cov}\{\textbf{h}_l[k]\} = \mathrm{Cov}\{\textbf{h}_l^H [k]\} = \textbf{I}_K$. By using Lemma 1, one can say $\mathbb{E}\big\{\textbf{h}_l^H [k] \textbf{A} \textbf{h}_l[k] \big\} = \mathrm{tr}(\textbf{A})$ which is equal to $M$. Therefore
\begin{equation}
S_k = \frac{1}{MK} \mathbb{E}_{s_k[i]} \big\{\big|s_k[i]\big|^2\big\} \left(M \sum_{l=0}^{L-1} d_l[k]\right)^2 = \frac{M\rho_f}{K}.
\label{eqn:15}
\end{equation}

Based on Lemma 1, a set of expectations necessary to calculate the noise power in (\ref{eqn:7}) is provided in the Appendix.

\noindent{\bf Proposition 2:} The effective noise power using CMFP is given as in (\ref{eqn:7}).\\
\noindent{\em Proof:\/} By an inspection of (\ref{eqn:5}), we notice that different terms are independent from each other. Therefore, by using the expectations (\ref{eqn:gmg4}) and (\ref{eqn:gmg7}), we can write the variance of the effective noise as
\ifCLASSOPTIONonecolumn
\begin{align*}
\mathrm{Var} \{n_k^\prime[i]\} = & \frac{1}{MK} \mathrm{Var} \bigg\{\sum_{l=0}^{L-1} \bigg(\textbf{F}_{(l,l)}[k,k] - \mathbb{E}\big\{\textbf{F}_{(l,l)}[k,k]\big\}\bigg) s_k[i]\bigg\} \\
& + \frac{1}{MK} \mathrm{Var}\bigg\{\sum_{\substack{b=1-L \\ b \neq 0}}^{L-1} \sum_{l=L_1}^{L_2} \textbf{F}_{(l,l-b)}[k,k] s_k[i-b]\bigg\} \\
& + \frac{1}{MK} \mathrm{Var}\bigg\{\sum_{\substack{q=1 \\ q \neq k}}^{K} \sum_{b=1-L}^{L-1} \sum_{l=L_1}^{L_2} \textbf{F}_{(l,l-b)}[k,q] s_q[i-b]\bigg\} + 1. \numberthis
\label{eqn:16}
\end{align*}
\else
\begin{align*}
\mathrm{Var} & \{n_k^\prime[i]\} = \\
& \frac{1}{MK} \mathrm{Var} \bigg\{\sum_{l=0}^{L-1} \bigg(\textbf{F}_{(l,l)}[k,k] - \mathbb{E}\big\{\textbf{F}_{(l,l)}[k,k]\big\}\bigg) s_k[i]\bigg\} \\
& + \frac{1}{MK} \mathrm{Var}\bigg\{\sum_{\substack{b=1-L \\ b \neq 0}}^{L-1} \sum_{l=L_1}^{L_2} \textbf{F}_{(l,l-b)}[k,k] s_k[i-b]\bigg\} \\
& + \frac{1}{MK} \mathrm{Var}\bigg\{\sum_{\substack{q=1 \\ q \neq k}}^{K} \sum_{b=1-L}^{L-1} \sum_{l=L_1}^{L_2} \textbf{F}_{(l,l-b)}[k,q] s_q[i-b]\bigg\}\\
& + 1. \numberthis
\label{eqn:16}
\end{align*}
\fi
Note that the means of IF, ISI, and MUI terms are zero. Considering the fact that $\mathrm{Var}\{s_k[i]\} = 1$ and the information symbols are independent from all other terms, after carefully reindexing (\ref{eqn:16}) and removing a number of redundant terms, we can rewrite the effective noise variance as
\ifCLASSOPTIONonecolumn
\begin{align*}
\mathrm{Var} \{n_k^\prime[i]\} = & \frac{\rho_f}{MK} \sum_{q=1}^K \sum_{b=1}^{L-1}\sum_{l=b}^{L-1}  \mathbb{E} \Big\{ {\bf F}_{(l,l-b)} [q,k] {\bf F}_{(l-b,l)}[k,q] \Big\} \\
& + \frac{\rho_f}{MK} \sum_{q=1}^K \sum_{b=1}^{L-1}\sum_{l=b}^{L-1} \mathbb{E} \Big\{ {\bf F}_{(l-b,l)}[q,k]{\bf F}_{(l,l-b)}[k,q]\Big\} \\
& + \frac{\rho_f}{MK} \sum_{q=1}^K \sum_{l=0}^{L-1} \mathbb{E} \Big\{ {\bf F}_{(l,l)} [q,k] {\bf F}_{(l,l)}[k,q] \Big\} - \frac{\rho_f}{MK}\sum_{l=0}^{L-1} \Big[ \mathbb{E} \Big\{ {\bf F}_{(l,l)}[k,k]\Big\} \Big]^2 + 1. \numberthis
\label{eqn:17}
\end{align*}
\else
\begin{align*}
& \mathrm{Var} \{n_k^\prime[i]\} = \\
& \frac{\rho_f}{MK} \sum_{q=1}^K \sum_{b=1}^{L-1}\sum_{l=b}^{L-1}  \mathbb{E} \Big\{ {\bf F}_{(l,l-b)} [q,k] {\bf F}_{(l-b,l)}[k,q] \Big\}\\
& + \frac{\rho_f}{MK} \sum_{q=1}^K \sum_{b=1}^{L-1}\sum_{l=b}^{L-1} \mathbb{E} \Big\{ {\bf F}_{(l-b,l)}[q,k]{\bf F}_{(l,l-b)}[k,q]\Big\} \\
& + \frac{\rho_f}{MK} \sum_{q=1}^K \sum_{l=0}^{L-1} \mathbb{E} \Big\{ {\bf F}_{(l,l)} [q,k] {\bf F}_{(l,l)}[k,q] \Big\}\\
& - \frac{\rho_f}{MK}\sum_{l=0}^{L-1} \Big[ \mathbb{E} \Big\{ {\bf F}_{(l,l)}[k,k]\Big\} \Big]^2 
+ 1. \numberthis
\label{eqn:17}
\end{align*}
\fi
Then, if we replace the expectations inside (\ref{eqn:17}) with their corresponding expressions in (\ref{eqn:gmg1})-(\ref{eqn:gmg7}) in the Appendix, the effective noise variance can be obtained as
\ifCLASSOPTIONonecolumn
\begin{align*}
\mathrm{Var}\{n_k^\prime[i]\}) =
& \frac{\mathrm{tr}(\textbf{A}^2)\rho_f}{MK} \sum_{q=1}^{K} \sum_{b=1}^{L-1} \sum_{l=b}^{L-1} \big(d_{l-b}[q]d_l[k] + d_l[q]d_{l-b}[k]\big) \\
& + \frac{\mathrm{tr}(\textbf{A}^2)\rho_f}{MK} \sum_{q=1}^{K} \sum_{l=0}^{L-1} d_l[k]d_l[q] + 1.\numberthis\label{eqn:gmg20}
\end{align*}
\else
\begin{align*}
& \mathrm{Var}\{n_k^\prime[i]\}) = \\
& \frac{\mathrm{tr}(\textbf{A}^2)\rho_f}{MK} \sum_{q=1}^{K} \sum_{b=1}^{L-1} \sum_{l=b}^{L-1} \big(d_{l-b}[q]d_l[k] + d_l[q]d_{l-b}[k]\big) \\
& + \frac{\mathrm{tr}(\textbf{A}^2)\rho_f}{MK} \sum_{q=1}^{K} \sum_{l=0}^{L-1} d_l[k]d_l[q] + 1.\numberthis\label{eqn:gmg20}
\end{align*}
\fi
One can recognize that (\ref{eqn:gmg20}) can be expressed as
\begin{equation}
\mathrm{Var}\{n_k^\prime[i]\} = \frac{{\rm tr}({\bf A}^2)\rho_f}{MK}\Big[\sum_{q=1}^K \Big( \sum_{l_1=0}^{L-1}d_{l_1}[q] \Big) \Big( \sum_{l_2=0}^{L-1}d_{l_2}[k] \Big) \Big] + 1
\label{eqn:gmg21}
\end{equation}
Then, by using (\ref{eqn:2}), (\ref{eqn:gmg21}) can be simplified as (\ref{eqn:7}).
\subsection{Conventional Precoders}

The zero-forcing precoder (ZFP) and regularized zero-forcing precoder (RZFP) are two well-known precoders in the massive MIMO field of study, see, e.g., \cite{C2}. A common theme among the precoders we will discuss in this paper is the fact that they are defined in the frequency domain and are translated into the time domain. Computations indicated that precoders defined in the time domain, such as those in \cite{C18}, do not perform as well as precoders in the frequency domain. The two precoders will be given as functions of the channel state matrix $\hat{\textbf{H}}_l$ as the following
\begin{itemize}
\item Zero-Forcing Precoder (ZFP): It forces the system to eliminate the interference and is given by
\begin{equation}
\textbf{W}_{\nu}^{\it ZFP} = a_W^{\it ZFP} \hat{\textbf{H}}_{\nu} \big(\hat{\textbf{H}}^H_{\nu} \hat{\textbf{H}}_{\nu}\big)^{-1},
\label{eqn:24}
\end{equation}
where $\hat{\textbf{H}}_{\nu} = \sum_{l=0}^{L-1} e^{-j2\pi \nu l/L} \hat{\textbf{H}}_l$ is the $N$-point ($N > L$) Fourier transform of the channel state matrix and $a_W^{\it ZFP}$ is a normalization factor for this precoder.
\item Regularized Zero-Focing Equalizer (RZFP): This precoder maximizes the power of the desired signal compared to the power of the noise and interference at the receiver. It is given by
\begin{equation}
\textbf{W}_{\nu}^{\it RZFP} = a_W^{\it RZFP} \hat{\textbf{H}}_{\nu} \big(\hat{\textbf{H}}^H_{\nu} \hat{\textbf{H}}_{\nu} + \beta_W \textbf{I}_K\big)^{-1},
\label{eqn:25}
\end{equation}
where $a_W^{\it RZFP}$ is also a power normalization factor and $\beta_W \in \mathbb{R}^+$ is a system parameter which depends on the signal-to-noise ratios (SNRs) and the path losses of the users.
\end{itemize}

Using these precoders, one can generate a new model for the transmit signal vector. Using a general notation of $\textbf{W}_{\nu}$ to indicate the precoder model,
the vector of the transmit signals can be obtained as
\begin{equation}
\textbf{x}[i] = \sum_{m=0}^{N-1} \textbf{W}_m \textbf{s}[i-m],
\label{eqn:27}
\end{equation}
where $\textbf{W}_m = \sum_{\nu=0}^{N-1} e^{j2\pi \nu m/N} \textbf{W}_{\nu}$ is the inverse Fourier transform of the precoder matrix. Note that (\ref{eqn:27}) represents the cyclic convolution and all the indices of equation defining the transmit signal are taken modulo $N$. Considering that (\ref{eqn:1}) still holds, the vector of the received signals at the users' site can be written as
\begin{equation}
\textbf{y}[i] = \sum_{m=0}^{N-1} \sum_{l=0}^{L-1} \hat{\textbf{H}}_l^H \textbf{W}_m \textbf{s}[i-l-m] + \textbf{n}[i].
\label{eqn:28}
\end{equation}
By defining the new super channel matrix as $\textbf{F}_{(l,m)} = \hat{\textbf{H}}_l^H \textbf{W}_m$ and considering the fact that (\ref{eqn:27}) represents circular convolution, one can rewrite (\ref{eqn:28}) as
\begin{equation}
\textbf{y}[i] = \sum_{m=1-N}^{0} \sum_{l=0}^{L-1} \textbf{F}_{(l,m)} \textbf{s}[i-l-m] + \textbf{n}[i].
\label{eqn:29}
\end{equation}
By changing the variable $m+l$ to $b$ and considering the fact that $\textbf{s}[i] = \sum_{q=1}^{K} \textbf{e}_q s_q[i]$, we can rewrite (\ref{eqn:29}) as
\begin{equation}
\textbf{y}[i] = \sum_{q=1}^{K} \sum_{b=1-N}^{L-1} \sum_{l=L_1}^{L_3} \textbf{F}_{(l,b-l)}[:,q] s_q[i-b] + \textbf{n}[i]
\label{eqn:30}
\end{equation}
where $\textbf{F}_{(l,b-l)}[:,q]$ is a column vector and where we defined $L_3 \triangleq \min(N-1+b, L-1)$. Note that the desired signal at the $k$-th user and time $i$ is given by
\begin{equation}
g_k[i] = \sum_{l=0}^{L-1} \mathbb{E}\big\{\textbf{F}_{(l,-l)}[k,k]\big\} s_k[i].
\label{eqn:31}
\end{equation}
Using the equations of the desired and received signal, one can express the system model in terms of desired signal and effective noise of the channel as
\begin{equation}
y_k[i] = g_k[i] + n_k^{\prime}[i],
\label{eqn:32}
\end{equation}
where $n_k^\prime[i]$ represents the effective noise and can be written as
\ifCLASSOPTIONonecolumn
\begin{align*}
n_k^\prime[i] = & \sum_{l=0}^{L-1} \bigg(\textbf{F}_{(l,-l)}[k,k] - \mathbb{E}\big\{\textbf{F}_{(l,-l)}[k,k]\big\}\bigg) s_k[i]
 + \sum_{\substack{b=1-N \\ b \neq 0}}^{L-1} \sum_{l=L_1}^{L_3} \textbf{F}_{(l,b-l)}[k,k] s_k[i-b] \\
& + \sum_{\substack{q=1 \\ q \neq k}}^{K} \sum_{b=1-N}^{L-1} \sum_{l=L_1}^{L_3} \textbf{F}_{(l,b-l)}[k,q] s_q[i-b] + n_k[i],\numberthis
\label{eqn:33}
\end{align*}
\else
\begin{align*}
n_k^\prime[i] & = \sum_{l=0}^{L-1} \bigg(\textbf{F}_{(l,-l)}[k,k] - \mathbb{E}\big\{\textbf{F}_{(l,-l)}[k,k]\big\}\bigg) s_k[i] \\
& + \sum_{\substack{b=1-N \\ b \neq 0}}^{L-1} \sum_{l=L_1}^{L_3} \textbf{F}_{(l,b-l)}[k,k] s_k[i-b] \\
& + \sum_{\substack{q=1 \\ q \neq k}}^{K} \sum_{b=1-N}^{L-1} \sum_{l=L_1}^{L_3} \textbf{F}_{(l,b-l)}[k,q] s_q[i-b] \\
& + n_k[i],\numberthis
\label{eqn:33}
\end{align*}
\fi
which again includes IF, ISI, MUI, and AWGN terms, respectively. The system model introduced in this section is used to run computations.
The achievable sum rate of the system can be obtained by considering the effect of noise into account as
\begin{equation}
R_{\it sum} = \frac{1}{2}\sum_{k=1}^{K} \log_2\big(1 + \frac{{\rm Var}\{g_k[i]\}}{\mathrm{Var}(n_k^\prime[i])}\big)
\label{eqn:33b}
\end{equation}
where it was proved that ${\rm Var}\{g_k[i]\}=\frac{M}{K}\rho_f$.
Equation (\ref{eqn:33b}) is the sum rate (information rate) for the precoders. It will also apply to the case of equalizers for the uplink channel as will be discussed in the next section. 


\section{Uplink Channel Analysis}
\label{sec:uplink}

Just like the downlink model, a frequency-selective multi-user MIMO (MU-MIMO) channel with $M$ base station antennas and $K$ single-antenna users is considered. Perfect knowledge of CSI is considered at the users' terminal in the uplink transmission. Since the correlation pattern is also considered in the uplink channel, the CSI matrix can be modeled as $\hat{\textbf{H}}_l = \textbf{A}^{1/2} \textbf{H}_l \textbf{D}_l^{1/2}$. Considering the uplink, users are supposed to transmit $\textbf{s}[i]$ through the channel. However, using equalizers in the frequency domain at the base station requires employment of cyclic prefix techniques.
In this work, the conventional cyclic prefix technique, where the last $T_C$ samples of a $T$-sample transmission block are added to the beginning of the block, is preferred over the zero-padding or the known symbol padding techniques due to its circular convolution property \cite{C19}.

In the same manner as it was introduced earlier in Section \ref{sec:downlink}, one can obtain the achievable sum rate of the uplink channel with an equalizer using (\ref{eqn:33b}). Note that $S_k$ is the power of the desired signal for the $k$-th user at the base station and $n_k^\prime[i]$ is the effective noise of the system for the $k$-th user at the base station at time $i$ considering the uplink channel.

\subsection{Channel Matched Filter Equalizer (CMFE)}

The cyclic prefix is designed in the way that the length of the added symbols are larger than the length of the channel (i.e., $T_c > L$).
We can write the received signal vector before applying the proper equalizer as
\begin{equation}
\textbf{r}[i] = \sum_{l=0}^{L-1} \hat{\textbf{H}}_l \textbf{x}[(i-l)\!\!\! \mod \, T] + \textbf{n}[i],
\label{eqn:a34}
\end{equation}
where $\textbf{x}[i]$ denotes the transmitted signal vector and consists of the transmitted symbols (e.g., $\sqrt{\rho_f} \textbf{s}[i]$) and the $T_c$ added symbols as the cyclic prefix.
In this section, the channel matched filter is considered as the {\em equalizer\/} in the system. For that reason we designate it as CMFE, as opposed to CMFP employed in the downlink channel. The following shows how CMFE will affect the unprocessed output signal of the channel.
\begin{equation}
\textbf{y}[i] = \frac{1}{\sqrt{MK}} \sum_{l=0}^{L-1} \hat{\textbf{H}}^H_l \textbf{r}[i+l],
\label{eqn:a35}
\end{equation}
where $\textbf{y}[i]$ is the vector of the received signal after the equalizer block. Substituting (\ref{eqn:a34}) in (\ref{eqn:a35}), one can obtain
\ifCLASSOPTIONonecolumn
\begin{equation}
\textbf{y}[i] = \sqrt{\frac{1}{MK}} \sum_{l=0}^{L-1} \sum_{l^\prime=0}^{L-1} \hat{\textbf{H}}^H_l \hat{\textbf{H}}_{l^\prime} \textbf{s}[(i-l^\prime+l)\!\!\! \mod \, T]
+ \frac{1}{\sqrt{MK}} \sum_{l=0}^{L-1} \hat{\textbf{H}}^H_l \textbf{n}[i+l].
\label{eqn:a36}
\end{equation}
\else
\begin{align}
\textbf{y}[i] & = \sqrt{\frac{1}{MK}} \sum_{l=0}^{L-1} \sum_{l^\prime=0}^{L-1} \hat{\textbf{H}}^H_l \hat{\textbf{H}}_{l^\prime} \textbf{s}[(i-l^\prime+l)\!\!\! \mod \, T] \nonumber\\
& + \frac{1}{\sqrt{MK}} \sum_{l=0}^{L-1} \hat{\textbf{H}}^H_l \textbf{n}[i+l].
\label{eqn:a36}
\end{align}
\fi
We can rewrite the received signal for each individual user at the base station as the following
\begin{equation}
y_k[i] = g_k[i] + n_k^\prime[i],
\label{eqn:a37}
\end{equation}
where $g_k[i]$ is the desired signal of the $k$-th user at the base station and can be written as
\begin{equation}
g_k[i] = \sqrt{\frac{1}{MK}} \bigg(\sum_{l=0}^{L-1} \textbf{F}_{(l,l)}[k,k] \bigg) s_k[i \!\!\! \mod \, T],
\label{eqn:a37-2}
\end{equation}
where $\textbf{F}_{(l,l^{\prime})} = \hat{\textbf{H}}_l^H \hat{\textbf{H}}_{l^{\prime}}$ is the super channel matrix.
The second is the effective noise of the system and, similar to the downlink channel, 
it can be written as
\ifCLASSOPTIONonecolumn
\begin{align*}
n_k^\prime[i] = & \sqrt{\frac{1}{MK}} \sum_{\substack{b=1-L \\ b \neq 0}}^{L-1}\sum_{l=L_1}^{L_2} \textbf{F}_{(l,l-b)}[k,k] s_k[(i-b) \!\!\! \mod \, T] \\
& + \sqrt{\frac{1}{MK}} \sum_{\substack{q=1 \\ q \neq k}}^{K} \sum_{b=1-L}^{L-1}\sum_{l=L_1}^{L_2} \textbf{F}_{(l,l-b)}[k,q] s_q[(i-b)\!\!\! \mod \, T]
    + \frac{1}{\sqrt{MK}} \sum_{l=0}^{L-1} \textbf{e}^T_k\hat{\textbf{H}}^H_l \textbf{n}[i+l]. \numberthis
\label{eqn:a38}
\end{align*}
\else
\begin{align*}
& n_k^\prime[i] = \\
& \sqrt{\frac{1}{MK}} \sum_{\substack{b=1-L \\ b \neq 0}}^{L-1}\sum_{l=L_1}^{L_2} \textbf{F}_{(l,l-b)}[k,k] s_k[(i-b) \!\!\! \mod \, T] \\
& + \sqrt{\frac{1}{MK}} \sum_{\substack{q=1 \\ q \neq k}}^{K} \sum_{b=1-L}^{L-1}\sum_{l=L_1}^{L_2} \textbf{F}_{(l,l-b)}[k,q] s_q[(i-b)\!\!\! \mod \, T] \\
& + \frac{1}{\sqrt{MK}} \sum_{l=0}^{L-1} \textbf{e}^T_k\hat{\textbf{H}}^H_l \textbf{n}[i+l]. \numberthis
\label{eqn:a38}
\end{align*}
\fi
As it can be seen, the effective noise term in the uplink is very similar to that of the downlink, except the difference of using CMFE makes in AWGN noise and perfect knowledge of CSI eliminates the IF term.

In order to determine the power of effective noise in the uplink channel, we need to determine the power of AWGN affected by CMFE, which is as the following
\ifCLASSOPTIONonecolumn
\begin{align*}
\mathrm{Var}(z_k[i]) & = \mathrm{Var}\bigg(\frac{1}{\sqrt{MK}} \sum_{l=0}^{L-1} \textbf{e}^T_k \hat{\textbf{H}}^H_l \textbf{n}[i+l]\bigg)
  = \frac{1}{MK} \mathbb{E} \bigg(\sum_{l=0}^{L-1} \textbf{e}^T_k \hat{\textbf{H}}^H_l \textbf{n}[i+l] \sum_{l^\prime=0}^{L-1} \textbf{n}^T[i+l^\prime] \hat{\textbf{H}}_{l^\prime} \textbf{e}_k\bigg) \\
& = \frac{1}{MK} \mathbb{E} \bigg(\sum_{l=0}^{L-1} \sum_{l^\prime=0}^{L-1} \textbf{e}^T_k \hat{\textbf{H}}^H_l \textbf{n}[i+l] \textbf{n}^T[i+l^\prime] \hat{\textbf{H}}_{l^\prime} \textbf{e}_k\bigg) \\
& = \frac{1}{MK} \mathbb{E}_{H_l}\! \bigg(\sum_{l=0}^{L-1} \sum_{l^\prime=0}^{L-1} \textbf{e}^T_k \hat{\textbf{H}}^H_l \mathbb{E}_{n[i]}\! \bigg(\textbf{n}[i+l] \textbf{n}^T[i+l^\prime]\! \bigg) \hat{\textbf{H}}_{l^\prime} \textbf{e}_k\bigg).
\numberthis
\end{align*}
\else
\begin{align*}
& \mathrm{Var}(z_k[i]) = \mathrm{Var}\bigg(\frac{1}{\sqrt{MK}} \sum_{l=0}^{L-1} \textbf{e}^T_k \hat{\textbf{H}}^H_l \textbf{n}[i+l]\bigg) \\
& \!\! = \frac{1}{MK} \mathbb{E} \bigg(\sum_{l=0}^{L-1} \textbf{e}^T_k \hat{\textbf{H}}^H_l \textbf{n}[i+l] \sum_{l^\prime=0}^{L-1} \textbf{n}^T[i+l^\prime] \hat{\textbf{H}}_{l^\prime} \textbf{e}_k\bigg) \\
& \!\! = \frac{1}{MK} \mathbb{E} \bigg(\sum_{l=0}^{L-1} \sum_{l^\prime=0}^{L-1} \textbf{e}^T_k \hat{\textbf{H}}^H_l \textbf{n}[i+l] \textbf{n}^T[i+l^\prime] \hat{\textbf{H}}_{l^\prime} \textbf{e}_k\bigg) \\
& \!\! = \frac{1}{MK} \mathbb{E}_{H_l}\! \bigg(\sum_{l=0}^{L-1} \sum_{l^\prime=0}^{L-1} \textbf{e}^T_k \hat{\textbf{H}}^H_l \mathbb{E}_{n[i]}\! \bigg(\textbf{n}[i+l] \textbf{n}^T[i+l^\prime]\! \bigg) \hat{\textbf{H}}_{l^\prime} \textbf{e}_k\bigg).
\numberthis
\end{align*}
\fi
By the nature of AWGN, if $l \neq l^\prime$, then $\mathbb{E}_{n[i]} \big(\textbf{n}[i+l] \textbf{n}^T[i+l^\prime] \big) = \textbf{0}_{M \times M}$. It will be equal to $\textbf{I}_M$ if $l = l^\prime$. Keeping this in mind, we can rewrite the power of the AWGN affected by CMFE as
\ifCLASSOPTIONonecolumn
\begin{align*}
\mathrm{Var}(z_k[i]) & =  \mathrm{Var}\bigg(\frac{1}{\sqrt{MK}} \sum_{l=0}^{L-1} \textbf{e}^T_k \hat{\textbf{H}}^H_l \textbf{n}[i+l]\bigg)
= \frac{1}{MK} \mathbb{E}_{H_l} \bigg(\sum_{l=0}^{L-1} \textbf{e}^T_k \hat{\textbf{H}}^H_l \hat{\textbf{H}}_{l^\prime} \textbf{e}_k\bigg) \\
& = \frac{1}{MK} \mathbb{E}_{H_l} \bigg(\sum_{l=0}^{L-1} \textbf{e}^T_k \textbf{D}^{1/2}_l \textbf{H}^H_l \textbf{A} \textbf{H}_l \textbf{D}^{1/2}_l \textbf{e}_k\bigg) = \frac{\mathrm{tr}(\textbf{A})}{MK}. \numberthis
\label{eqn:z11}
\end{align*}
\else
\begin{align*}
\mathrm{Var}(z_k[i]) &= \mathrm{Var}\bigg(\frac{1}{\sqrt{MK}} \sum_{l=0}^{L-1} \textbf{e}^T_k \hat{\textbf{H}}^H_l \textbf{n}[i+l]\bigg)\\
& = \frac{1}{MK} \mathbb{E}_{H_l} \bigg(\sum_{l=0}^{L-1} \textbf{e}^T_k \hat{\textbf{H}}^H_l \hat{\textbf{H}}_{l^\prime} \textbf{e}_k\bigg) \\
& = \frac{1}{MK} \mathbb{E}_{H_l} \bigg(\sum_{l=0}^{L-1} \textbf{e}^T_k \textbf{D}^{1/2}_l \textbf{H}^H_l \textbf{A} \textbf{H}_l \textbf{D}^{1/2}_l \textbf{e}_k\bigg) \\
& = \frac{\mathrm{tr}(\textbf{A})}{MK}.
\numberthis
\label{eqn:z11}
\end{align*}
\fi
Therefore, using the power of AWGN affected by CMFE in (\ref{eqn:z11}) and the power of interference terms calculated in \cite{C25} neglecting the IF term, one can obtain the power of the effective noise in the uplink channel as
\begin{equation}
\mathrm{Var}(n_k^\prime[i]) = \frac{\mathrm{tr}(\textbf{A}^2)}{M}\bigg(1 - \sum_{l=0}^{L-1} \frac{d_l^2[k]}{K}\bigg)\rho_f + \frac{\mathrm{tr}(\textbf{A})}{MK},
\end{equation}
where $\mathrm{tr}(\textbf{A}) = M$ for the correlation patterns we consider in this work.

Based on what is stated in \cite{C25}, the power of the desired signal can be obtained as $\frac{\rho_f}{MK} \big(\mathrm{tr}(\textbf{A}^2)\sum_{l=0}^{L-1}d_l^2[k] + \mathrm{tr}^2(\textbf{A})(\sum_{l=0}^{L-1}d_l[k])^2\big)$. Note that $\sum_{l=0}^{L-1}d_l[k]=1$
One can obtain the achievable sum rate as
\begin{equation}
R_{sum} = \frac{K}{2} \log_2 \bigg(1 + \frac{\mathrm{tr}(\textbf{A}^2)\sum_{l=0}^{L-1}d_l^2[k]\rho_f + \mathrm{tr}^2(\textbf{A})\rho_f}{\mathrm{tr}(\textbf{A}^2)(K - \sum_{l=0}^{L-1}d_l^2[k])\rho_f + M}\bigg).
\end{equation}

\subsection{Conventional Equalizers}

Clearly, CMFE is a special case of implementing equalizers in the system. Using the general notation of $\textbf{Q}_{\nu}$ to denote the equalizer equation in the frequency domain, one can obtain the received signal as
\begin{equation}
\textbf{y}[f] = \textbf{Q}_{\nu} \textbf{r}[f],
\label{eqn:37}
\end{equation}
where $\textbf{r}[f] = \sum_{i=0}^{T-1} e^{-j2\pi fi/T} \textbf{r}[i]$ and $\textbf{y}[f] = \sum_{i=0}^{T-1} e^{-j2\pi fi/T} \textbf{y}[i]$ are the Fourier transforms of the unprocessed signal vector and the received signal vector, respectively.

Selecting the equalizer scheme has been the subject of study in MIMO channel as in \cite{C18}. Two well-known equalizers are considered in this work:
\begin{itemize}
\item Zero-Forcing Equalizer (ZFE): As it was introduced earlier as a precoder, ZFE can bring down the ISI to zero. This equalizer is given by the following
\begin{equation}
\textbf{Q}_{\nu}^{\it ZFE} = a_Q^{\it ZFE} \big(\textbf{G}_{\nu} \textbf{G}_{\nu}^H\big)^{-1} \textbf{G}_{\nu},
\label{eqn:38}
\end{equation}
where $a_Q^{\it ZFE}$ is the normalization factor, $\textbf{G}_\nu = \sum_{l=0}^{L-1} e^{-j2\pi \nu l/L} \textbf{G}_l$ is the Fourier transform of the composite channel matrix, and ${\bf G}_l=\hat {\bf H}_l^H$.
\item Minimum Mean Square Error Equalizer (MMSEE): As it is known in the downlink channel as RZFP, MMSEE is able to maximize the power of received signal compared to the noise and interference. This equalizer can be obtained as
\begin{equation}
\textbf{Q}_{\nu}^{\it MMSEE} = a_Q^{\it MMSEE} \big(\textbf{G}_{\nu} \textbf{G}_{\nu}^H + \beta_Q \textbf{I}_K\big)^{-1} \textbf{G}_{\nu},
\label{eqn:39}
\end{equation}
where $a_Q^{\it MMSEE}$ is also a normalization factor and $\beta_Q \in \mathbb{R}^+$ is a system parameter just like the one in RZFP, which depends on the SNRs and the path losses of the users. By inspecting the similarities in the definition of MMSEE and RZFP, it can be concluded that MMSEE in the uplink performs the same functionality as RZFP in the downlink.
\end{itemize}

Considering the inverse Fourier transform of the equalizer matrix $\textbf{Q}_m = \sum_{\nu=0}^{N-1} e^{j2\pi \nu m/N} \textbf{Q}_{\nu}$, (\ref{eqn:37}) can be rewritten in the time domain as
\ifCLASSOPTIONonecolumn
\begin{equation}
\textbf{y}[i] = \sum_{m=0}^{N-1} \sum_{l=0}^{L-1} \textbf{Q}_m \hat{\textbf{H}}_l \textbf{s}[(i-l-m)\!\!\! \mod \, T] + \sum_{m=0}^{N-1} \textbf{Q}_m \textbf{n}[i-m].
\label{eqn:40}
\end{equation}
\else
\begin{align}
\textbf{y}[i] = & \sum_{m=0}^{N-1} \sum_{l=0}^{L-1} \textbf{Q}_m \hat{\textbf{H}}_l \textbf{s}[(i-l-m)\!\!\! \mod \, T] \nonumber\\
                + & \sum_{m=0}^{N-1} \textbf{Q}_m \textbf{n}[i-m].
\label{eqn:40}
\end{align}
\fi

Note that using an equalizer in the channel, the super CSI matrix can be obtained by $\textbf{F}_{(m,l)} = \textbf{Q}_m \hat{\textbf{H}}_l$. Similar to the downlink channel, one can rewrite the vector of the received signal using the desired signal and the effective noise terms
\begin{equation}
y_k[i] = g_k[i] + n^{\prime}_k[i],
\label{eqn:41}
\end{equation}
where $g_k[i]$ is the desired signal at the $k$-th user and time $i$
\begin{equation}
g_k[i] = \sum_{l=0}^{L-1} \textbf{F}_{(-l,l)}[k,k] s_k[i].
\label{eqn:42}
\end{equation}
Note that the channel length is smaller than the frequency range (i.e., $N > L$). One can obtain the effective noise of the channel at the $k$-th user as the following
\ifCLASSOPTIONonecolumn
\begin{align}
n_k^\prime[i] =
& \sum_{\substack{b=1-N \\ b \neq 0}}^{L-1} \sum_{l=L_1}^{L_3} \textbf{F}_{(b-l,l)}[k,k] s_k[i-b] \nonumber\\
+ & \sum_{\substack{q=1 \\ q \neq k}}^{K} \sum_{b=1-N}^{L-1} \sum_{l=L_1}^{L_3} \textbf{F}_{(b-l,l)}[k,q] s_q[i-b] + \sum_{q=1}^{K} \sum_{m=0}^{N-1} \textbf{Q}_m[k] n_q[i-m],
\label{eqn:33c}
\end{align}
\else
\begin{align}
n_k^\prime[i] =
& \sum_{\substack{b=1-N \\ b \neq 0}}^{L-1} \sum_{l=L_1}^{L_3} \textbf{F}_{(b-l,l)}[k,k] s_k[i-b] \nonumber\\
 + & \sum_{\substack{q=1 \\ q \neq k}}^{K} \sum_{b=1-N}^{L-1} \sum_{l=L_1}^{L_3} \textbf{F}_{(b-l,l)}[k,q] s_q[i-b] \nonumber\\
 + & \sum_{q=1}^{K} \sum_{m=0}^{N-1} \textbf{Q}_m[k] n_q[i-m],
\label{eqn:33c}
\end{align}
\fi
where it consists of ISI, MUI, and independent noise terms, respectively, missing the IF term that exists in the downlink channel.
\subsection{Importance of Full CSI at the Base Station}\label{sec:importanceofcsi}
Note that, since this is uplink transmission, the full CSI is available at the base station. This enables expression of the desired response $g_k[i]$ as in (\ref{eqn:a37-2}) in the case of CFME and as (\ref{eqn:42}) in the case of MMSEE and ZFE. If the full CSI were not available, for example, as it would be in the case of downlink transmission, $g_k[i]$ for CMFE would have to be expressed as
\begin{equation}
g_k[i] = \sqrt{\frac{1}{MK}} \bigg(\sum_{l=0}^{L-1} \mathbb{E}\big\{\textbf{F}_{(l,l)}[k,k]\big\} \bigg) s_k[i \!\!\! \mod \, T],
\end{equation}
instead of (\ref{eqn:a37}), resulting in an additional interference term (IF)
\begin{equation}
\sum_{l=0}^{L-1} \bigg(\textbf{F}_{(-l,l)}[k,k] - \mathbb{E}\big\{\textbf{F}_{(-l,l)}[k,k]\big\}\bigg) s_k[i] .
\end{equation}
in (\ref{eqn:a38}), and a similar condition would hold in the case of MMSEE and ZFE in (\ref{eqn:42}) and (\ref{eqn:33c}), as was the case in (\ref{eqn:4}) and (\ref{eqn:5}). The IF interference represents the instantaneous wander of ${\bf F}_{(l,l)}[k,k]$ around its mean. In general, it causes a large degree of interference. Its absence in the uplink is a significant advantage.


\section{Correlation Patterns}
\label{sec:correlation}

A recent paper discusses spatially correlated antennas for massive MIMO systems in detail \cite{SBH19}. The paper's main conclusion is that the acquisition and utilization of spatial correlation information will be very important in wireless systems of the future, to take the spectral efficiency to the next level. This paper provides a number of observations along this basic idea. Recognizing the importance of this subject earlier, a number of papers on the characterization of spatial correlation for Massive MIMO systems and potential methods to exploit this phenomenon have recently appeared \cite{NKDA16,AA18,DL19,DZC19}.

As experimental evidence for spatial correlation, an S-parameter-based formulation shows that when only two monopoles are considered, the coefficient varies from 0.8 to about 0.2 when the antenna separation is about 0.05 to 0.2 times the wavelength \cite{C14}. A measurement study shows that, depending on propagation conditions, spatial correlation can remain significantly higher than 0.9 for a wide range of antenna separation values \cite{C15}. A range of average $a$ values from 0.4 to 0.7 for antenna separations at approximately 0.25 to 0.5 wavelength was reported in \cite{C17}.

In this section, two of the most well-known correlation patterns, the exponential correlation pattern and the Bessel function correlation pattern, are introduced and elaborated on. Considering the structure of the two patterns, the correlation matrix $\textbf{A}$ will remain symmetric.

\subsection{Exponential Correlation Model}

We will consider a linear antenna array for this correlation model. Assume antenna $m_i$ is in position $i$ and antenna $m_j$ is in position $j$. It is reasonable to expect, and is evidenced in the literature, that the effect of these antennas on one another should be related to $|i-j|$ considering the dependency of the channel to be among the antennas. A basic correlation factor $0 < \alpha < 1$ is considered, which shows the effects of antennas with respect to $|i-j|$ on each other, where $\alpha$ is a real number. One can obtain a correlation matrix based on this model in which the elements are $[{\bf A}]_{i,j} = \alpha^{|i-j|d}$ where $d$ is the distance between the antennas, normalized with respect to the wavelength and is considered uniform between antenna elements. The matrix in this model is known as the exponential correlation matrix. This model is commonly employed when spatial correlation is considered for MIMO or spatial diversity channels, see, e.g., \cite{C10, C13}. The correlation coefficient increases as the separation between antennas decreases.

\subsection{Bessel Correlation Model}

Another correlation model in MIMO systems is introduced in \cite{AK02}. We refer to this model as the Bessel function correlation model. The $(i,j)$-th element of the new correlation matrix can be obtained by
\begin{equation}
[\textbf{A}]_{i,j} = \frac{I_{0} \big( \sqrt{\eta^2 - 4\pi^2d_{i,j}^2 + {\hat\imath}4\pi\eta\sin(\mu)d_{i,j}} \big)}{I_{0} (\eta)},
\label{eqn:34}
\end{equation}
where $0 < \eta < \infty$ controls the width of the AOA, $\mu \in [-\pi, \pi)$ is the mean direction of the AOA, $I_0$ is the zero-order modified Bessel function, ${\hat\imath} = \sqrt{-1}$ and $d_{i,j}$ is the distance between $j$-th and $i$-th antenna normalized with respect to the wavelength \cite{AK02}. We consider a linear array for the antenna elements at the base station in this model as well. Therefore, we can denote the distance between $i$-th and $j$-th antenna as
\begin{equation}
d_{i,j} = |i-j|d,
\label{eqn:35}
\end{equation}
where $d$ is the distance between adjacent antenna elements, normalized with respect to the wavelength.

In the model in (\ref{eqn:34}), $\mu=0$ corresponds to the perpendicular direction from the linear antenna array (broadside direction), while $\mu = \pm \pi/2$ correspond to the parallel direction to the linear antenna array (end-fire or inline direction). Note that, for all values of $\mu$, since $I_0(x)$ is a uniformly increasing function of $x$, $[{\bf A}]_{i,i}=1$, $[{\bf A}]_{i,j}<[{\bf A}]_{i,k}$ for $k>j$, and ${\bf A}$ will be symmetric. Our focus with this model in this paper is on the spatial multiplexing of closely spaced users positioned in the same beam of the massive array.

The parameter $\eta$ decreases the achievable rate of the system dramatically by changing from $0$ (isotropic scattering) to $\infty$ (extremely non-isotropic). The isotropic scattering model, also known as the Clarke's model, corresponds to the uniform distribution for the AOA \cite{clarke68,jakes74}. However, empirical measurements have shown that the AOA distribution of waves impinging on the user is more likely to be nonuniform \cite{AK02}. It is shown in \cite{ABK02} that values of $\eta$ show a very large variation, pointing out to the substantial change in the achievable rate with this fact of propagation.

The distance properties for ULA are straightforward and are discussed above. We will describe the extension of the distance properties to UPA in Section~\ref{sec:upa}.


\section{Computation Results}
\label{sec:simulations}

For the purpose of computations, PDP is considered to be exponential with $L=4$ and $d_l[k] = \frac{e^{-\theta_kl}}{\sum_{i=0}^{3} e^{-\theta_ki}}$, $l = \{0, \cdots, 3\}$, where $\theta_k = \frac{K-1}{5}$, $k = \{1, \cdots, K\}$. Note that the achievable sum rate is invariant of the channel PDP. Hence, any other PDP which satisfies (\ref{eqn:2}) can also be considered.
\subsection{Uniform Linear Array (ULA)}\label{sec:ula}
We will first discuss the case of a ULA at the base station (BS), see, e.g., \cite[Fig.~1]{WBL17}. We will consider both a downlink and an uplink channel. We assume that there are $M$ antennas placed on a line, at a distance of $d$ from one another. In our simulations, we will assume that $d$ is equal to half of the wavelength of transmission. We will consider both the exponential and the Bessel model with a set of values for $\alpha$ in the exponential channel model and $\eta$ in the Bessel channel model. In the Bessel model, we will assume arrival from the broadside $\mu=0$, as well as other angles, i.e., $\pi/4$, $\pi/3$, and $\pi/2$.
\subsubsection{Downlink Channel: Exponential Correlation Pattern}

\begin{figure}[!t]
\centering
\ifCLASSOPTIONonecolumn
\includegraphics[width=11.5cm]{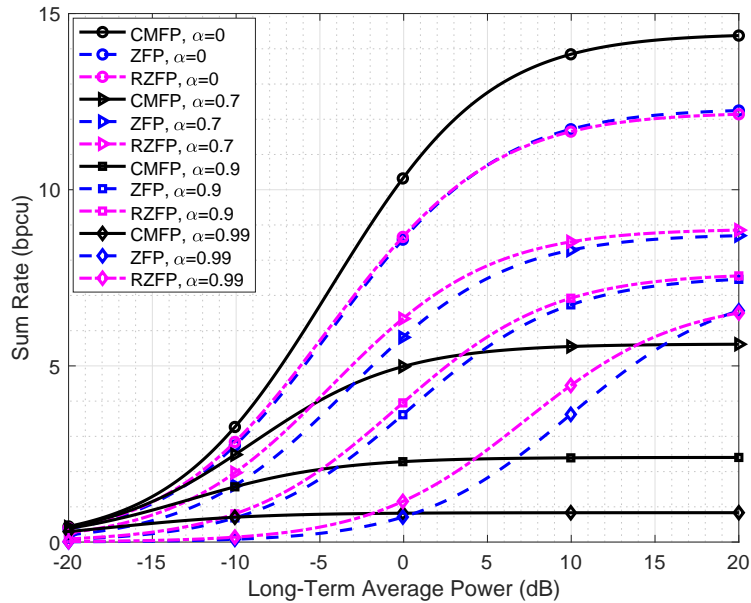}
\else
\includegraphics[width=6.5cm]{Figures/dl_1D_exp}
\fi
\vspace{-3.5mm}
\caption{Achievable sum rates of the three precoders in a downlink channel under the exponential correlation pattern with correlation parameter $\alpha \in \{0, 0.7, 0.9, 0.99\}$. The number of antennas at the base station is $M=64$, the number of users is $K=10$, the number of taps used in the computations is $L=4$, and the block length in the time domain is $T=100$. The distance between adjacent antenna elements is half the wavelength.}
\label{fig:dl_exp}
\end{figure}

Fig.~\ref{fig:dl_exp} shows a comparison of the achievable sum rate in a downlink channel, where the antenna elements are correlated with an exponential correlation pattern with $\alpha \in \{0, 0.7, 0.9, 0.99\}$. The parameters of the antenna array are given in the figure caption. We observe that the sum rate performance for the spatially uncorrelated channel $\alpha=0$ is such that CMFP outperforms the other two encoders, as expected. This changes for $\alpha=0.7$, however, and RZFP performs better than CMFP for average power around -5 dB and higher, while ZFP performs better than CMFP for average power -2 dB and higher. As $\alpha$ increases, the amount with which ZFP and RZFP outperform CMFP increases with an advantage for RZFP in the range of power in Fig.~\ref{fig:dl_exp}. Observe that as $\alpha$ increases from 0.7, the performance of RZFP and ZFP is a significant multiple  of that of CMFP at full power in Fig.~\ref{fig:dl_exp}. We note that, in order to run the computations for RZFP, we set the value for the system parameter $\beta_W$ in the manner to maximize the sum rate of the users. How to find such optimal $\beta_W$ is described in \cite{C2}, \cite{C16}.

\subsubsection{Downlink Channel: Bessel Correlation Matrix}
\begin{figure}[!t]
\centering
\ifCLASSOPTIONonecolumn
\includegraphics[width=11.5cm]{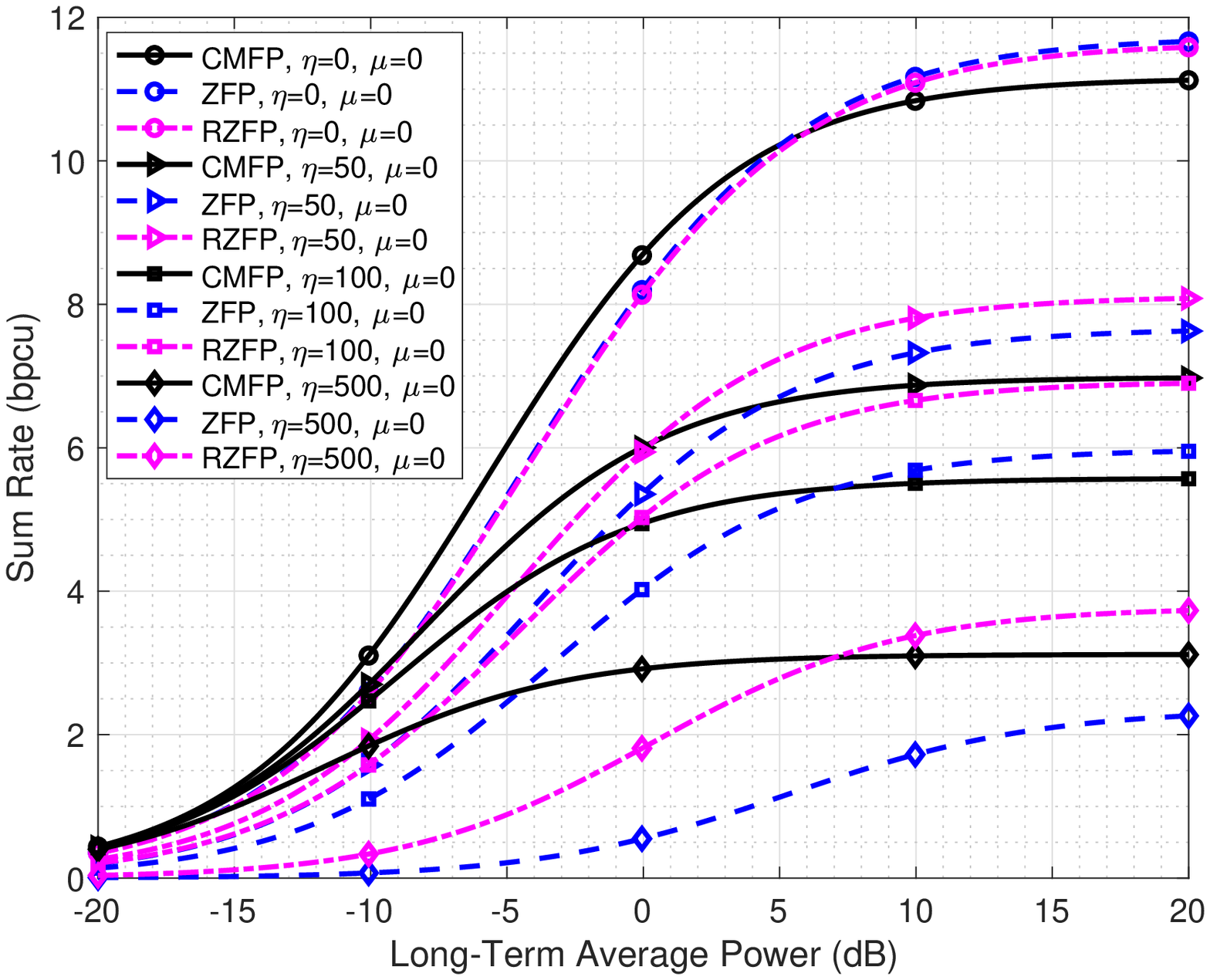}
\else
\includegraphics[width=6.5cm]{Figures/dl_1D_bsl}
\fi
\vspace{-3.5mm}
\caption{Achievable sum rates of the three precoders in a downlink channel under the Bessel correlation pattern with correlation parameters $\eta \in \{0, 50, 100, 500\}$ and $\mu = 0$. System parameters are $M=64$, $K=10$, $L=4$, $T=100$. The distance between adjacent antenna elements is half the wavelength.}
\label{fig:dl_bsl}
\centering
\ifCLASSOPTIONonecolumn
\includegraphics[width=11.5cm]{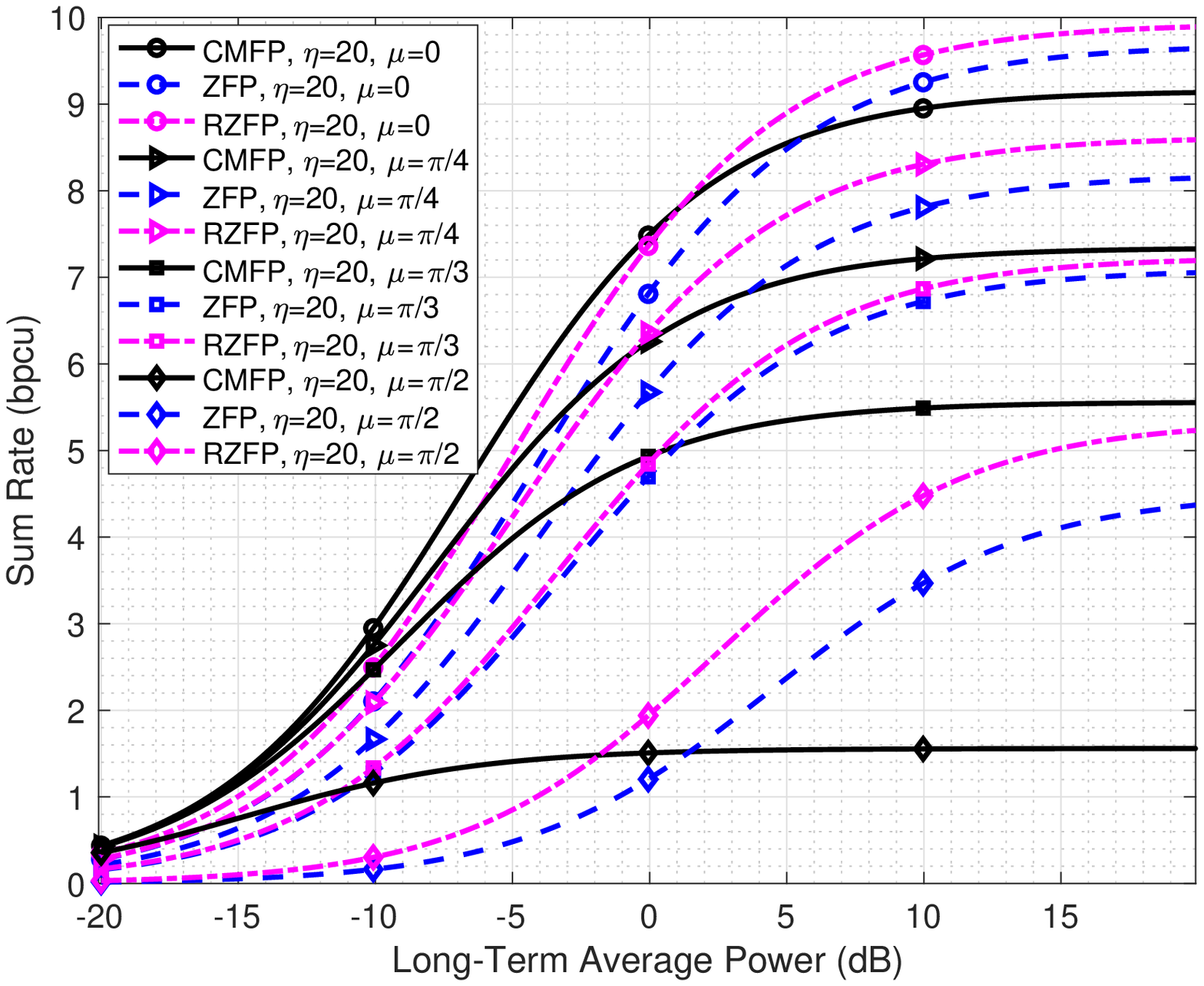}
\else
\includegraphics[width=6.5cm]{Figures/dl_1D_mu}
\fi
\vspace{-3.5mm}
\caption{Achievable sum rates of the three precoders in a downlink channel under the Bessel correlation pattern with correlation parameters $\mu \in \{0, \pi/4, \pi/3, \pi/2\}$ and $\eta = 20$. System parameters are $M=64$, $K=10$, $L=4$, $T=100$. The distance between adjacent antenna elements is half the wavelength.}
\label{fig:dl_mu}
\end{figure}

In a similar fashion, Fig.~\ref{fig:dl_bsl} shows a comparison of the achievable sum rate in a downlink channel, where the antenna elements are correlated with a Bessel correlation pattern with $\eta \in \{0, 50, 100, 500\}$ and $\mu = 0$. Recall that $\eta=0$ corresponds to an isotropic arrival, while increasing $\eta$ values correspond to increasing degrees of nonisotropic arrivals. The parameters of the antenna array are the same as in Fig.~\ref{fig:dl_exp} and are also given in the figure caption. We observe that the sum rate performance for the spatially uncorrelated channel $\eta=0$ is such that RZFP and ZFP outperform CMFP for power values beyond 6 dB for $\alpha = 0$, 0.7 and 0.9. As $\eta$ increases, RZFP outperforms CMFP beyond power levels of 0, 0, and 7 dB for $\eta = 50, 100,$ and $500$, respectively.

Fig.~\ref{fig:dl_mu} shows the achievable rates for $\eta=20$ and $\mu \in \{0, \pi/4, \pi/3, \pi/2\}$. Note that RZFP performs much better than CMFP as $\mu$ increases. Even the performance of ZFP becomes better than CMFP for larger values of $\mu$. Also, note that the end-fire or inline direction corresponding to $\pi/2$ has a nonzero sum rate value because of the presence of nonzero correlation among antenna elements.

We note that in Fig.~\ref{fig:dl_exp}--\ref{fig:dl_mu}, CMFP performs the best at low power formulas. As a result, it can be recommended to switch the precoder type based on long-term average power.
\subsubsection{Uplink Channel: Exponential Correlation Pattern}
\begin{figure}[!t]
\centering
\ifCLASSOPTIONonecolumn
\includegraphics[width=11.5cm]{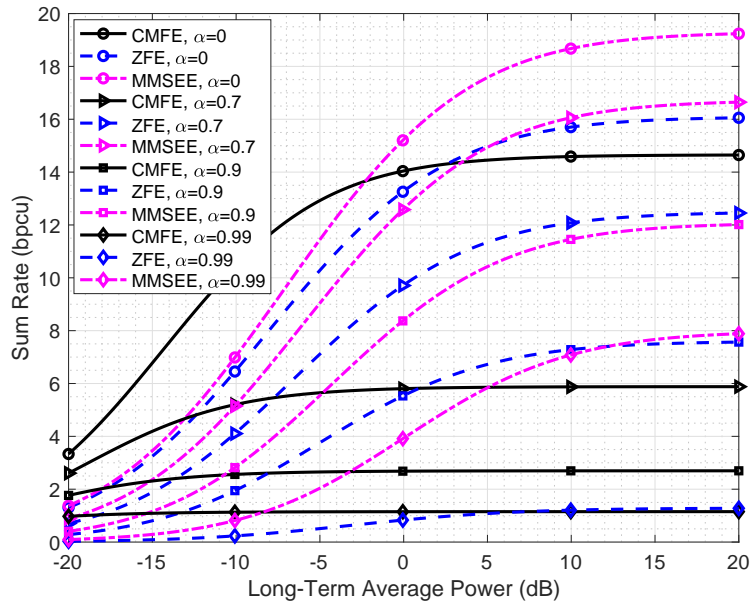}
\else
\includegraphics[width=6.5cm]{Figures/ul_1D_exp}
\fi
\vspace{-3.5mm}
\caption{Achievable sum rates of the three equalizers in the uplink channel under the exponential correlation pattern with correlation parameter $\alpha \in \{0, 0.7, 0.9, 0.99\}$. System parameters are $M=64$, $K=10$, $L=4$, $T=100$, $N=20$, $T_c=20$. The distance between adjacent antenna elements is half the wavelength.}
\label{fig:ul_exp}
\end{figure}

Fig.~\ref{fig:ul_exp} shows a comparison of the achievable sum rate in an uplink channel, where the antenna elements are correlated with an exponential correlation pattern with $\alpha \in \{0, 0.7, 0.9, 0.99\}$. In order to obtain $\beta_Q$ to run the computations for MMSEE, we used the same procedure to obtain $\beta_W$ for RZFP, in the way to maximize the users' sum rate with respect to the received SNR, see, i.e., \cite{C2} and \cite{C16}.

We now observe a very different behavior. The performance of MMSEE is much better than CMFE for all values of $\alpha$, including $\alpha=0$ at high average power values. Note that the gains with MMSEE at full power are a significant multiple of those of CMFE. This is because of the fact that since BS has access to CSI, it can force IF to be equal to zero, as described in Section~\ref{sec:importanceofcsi}. In addition, the performance of ZFE can also be better than CMFE, although not as much as MMSEE.
\subsubsection{Uplink Channel: Bessel Correlation Pattern}

\begin{figure}[!t]
\centering
\ifCLASSOPTIONonecolumn
\includegraphics[width=11.5cm]{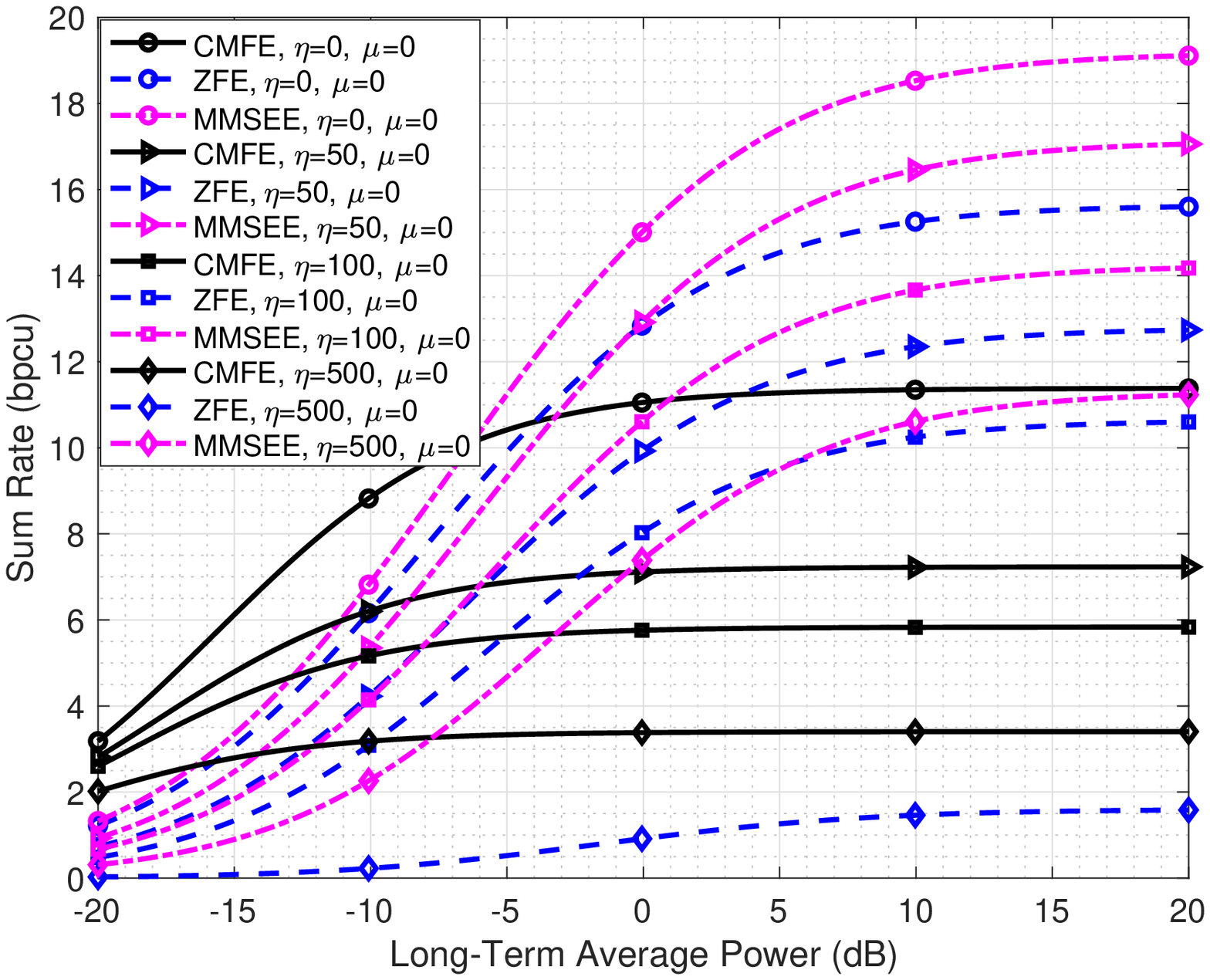}
\else
\includegraphics[width=6.5cm]{Figures/ul_1D_bsl}
\fi
\vspace{-3.5mm}
\caption{Achievable sum rates of the three equalizers in the uplink channel under the Bessel correlation pattern with correlation parameters $\eta \in \{0, 50, 100, 500\}$ and $\mu = 0$. System parameters are $M=64$, $K=10$, $L=4$, $T=100$, $N=20$, $T_c=20$. The distance between adjacent antenna elements is half the wavelength.}
\label{fig:ul_bsl}
\centering
\ifCLASSOPTIONonecolumn
\includegraphics[width=11.5cm]{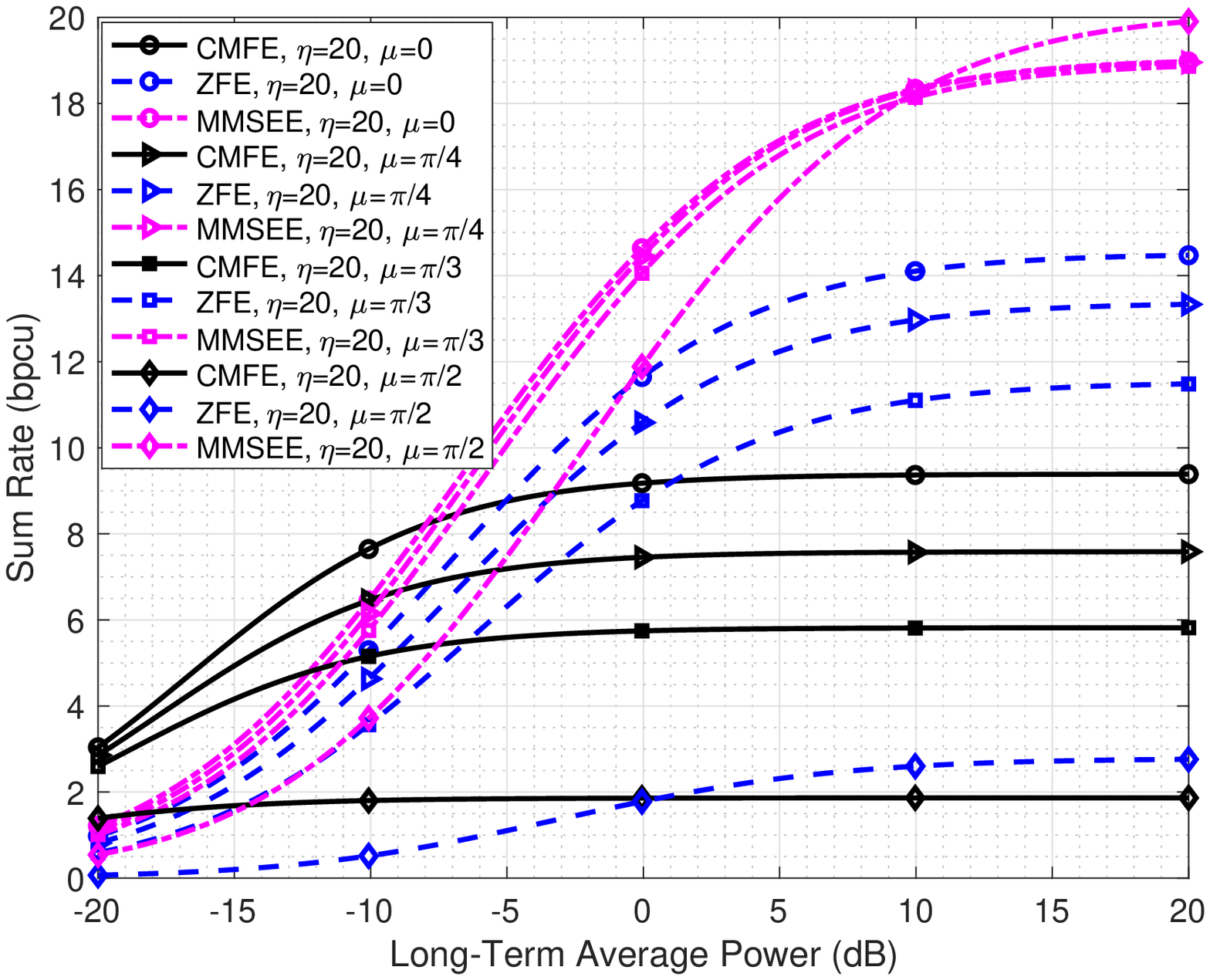}
\else
\includegraphics[width=6.5cm]{Figures/ul_1D_mu}
\fi
\vspace{-3.5mm}
\caption{Achievable sum rates of the three equalizers in the uplink channel under the Bessel correlation pattern with correlation parameters $\mu \in \{0, \pi/4, \pi/3, \pi/2\}$ and $\eta = 20$. System parameters are $M=64$, $K=10$, $L=4$, $T=100$, $N=20$, $T_c=20$. The distance between adjacent antenna elements is half the wavelength.}
\label{fig:ul_mu}
\end{figure}

Fig.~\ref{fig:ul_bsl} shows the comparison of the achievable sum rate in an uplink channel, where the antenna elements are correlated with a Bessel correlation pattern with $\eta \in \{0, 50, 100, 500\}$ and $\mu = 0$. We observe that the sum rate performance for all values of $\eta$ is such that MMSEE and ZFE outperform CMFE at sufficiently large power. In this case, the performance gain with MMSEE against CMFE is even larger than the exponential channel in Fig.~\ref{fig:ul_exp}.

Fig.~\ref{fig:ul_mu} shows the comparison of the achievable sum rate in an uplink channel, where the antenna elements are correlated with a Bessel correlation pattern with $\mu \in \{0, \pi/4, \pi/3, \pi/2\}$ and $\eta = 20$. Clearly, the performance of MMSEE is much better than that of CMFE for large values of the average power, where the crossover point is much less than all the other cases we have investigated in this paper so far. The performance of ZFE is better than CMFE, but it is not as significantly better than CMFE as MMSEE is. Both for Fig.~\ref{fig:ul_bsl} and Fig.~\ref{fig:ul_mu}, the improvement in performance with MMSEE at full power is significant and, increasingly with $\mu$, even a substantial multiple of that of CMFE.
\subsection{Uniform Planar Array (UPA)}\label{sec:upa}
We will now discuss the case of a UPA at the BS, see, e.g., \cite[Fig.~2]{WBL17}. Our treatment of this topic will follow that of Section~\ref{sec:ula}. We will consider both a downlink and an uplink channel. We assume that there are a total of $M$ antennas placed on a rectangle, at horizontal and vertical distances of $d_x=d_y=d$. In our simulations, we will assume that $d$ is equal to half of the wavelength of transmission. As in Section~\ref{sec:ula}, we will consider both the exponential and the Bessel correlation model with a set of values for $\alpha$ in the exponential channel model and $\eta$ in the Bessel channel model. In the Bessel model, we will assume arrival from the broadside $\mu=0$, as well as other angles, i.e., $\pi/4$, $\pi/3$, and $\pi/2$. Again, recall that $\eta=0$ corresponds to an isotropic arrival, while increasing $\eta$ values correspond to increasing degrees of nonisotropic arrivals.

We assume that UPA is located in the $x$-$z$ plane while the user is located in the $x$-$y$ plane. In a UPA, each antenna element undergoes spatial correlation in both spatial dimensions. Therefore, in a UPA, the effect of correlation will be higher. As a result, it can be expected that the sum rate will in general be lower as compared to a ULA. On the other hand, the performance of RZFP and ZFP can be expected to be better than CMFP, more than the case for ULA. Likewise, the performance of MMSEE and ZFE can be expected to be better than CMFE, again, more than the case for ULA.

The total number of antennas can be written as $M = M_x \times M_z$ where $M_x$ and $M_z$ indicate the number of antennas at each row and column, respectively. Clearly, the distance matrix for each row of UPA has the same structure as that for ULA. Hence, the distance matrix has the same linear property (increasing linearly as the indices increase) for the antenna elements in the same row or column. For calculating the distance between antenna elements, one only needs to add an offset, indicating the difference between row indices and column indices. Assume that the antenna elements $m$ are numbered starting at $m=0$ with the element in the first row and first column, and incrementing by 1 at each element for the first row, and then starting at $M_x$ for the second row and then incrementing by 1 at each element for the second row, and continuing this way until the last row and last column is reached which will have the value of $M - 1$. We define $r_m$ as the position of antenna element $m$ in rows of the structure in terms of $d$. Therefore, $r_m$ will be an integer running from 0 to $M_x-1$. The value of $r_m$ is the closest integer less than or equal to $m/M_x$, i.e.,
\begin{equation}
r_m = \Big\lfloor \frac{m}{M_x} \Big\rfloor .
\end{equation}
In a similar manner, one can obtain the position of antenna element $m$ in columns as
\begin{equation}
c_m = m - r_m M_x
\end{equation}
where $c_m$ ranges from 0 to $M_z-1$. Considering the different positioning of antenna elements, one can obtain a general formula to calculate the distance between antenna element $i$ and antenna element $j$ as
\begin{equation}
d_{ij} = d \sqrt{\zeta^2 + |c_i - c_j|^2}
\end{equation}
where $\zeta = | r_i - r_j |$ working as the offset and $d$ is the normalized distance between adjacent antennas.
\subsubsection{Downlink Channel: Exponential Correlation Pattern}
\begin{figure}[!t]
\centering
\ifCLASSOPTIONonecolumn
\includegraphics[width=11.5cm]{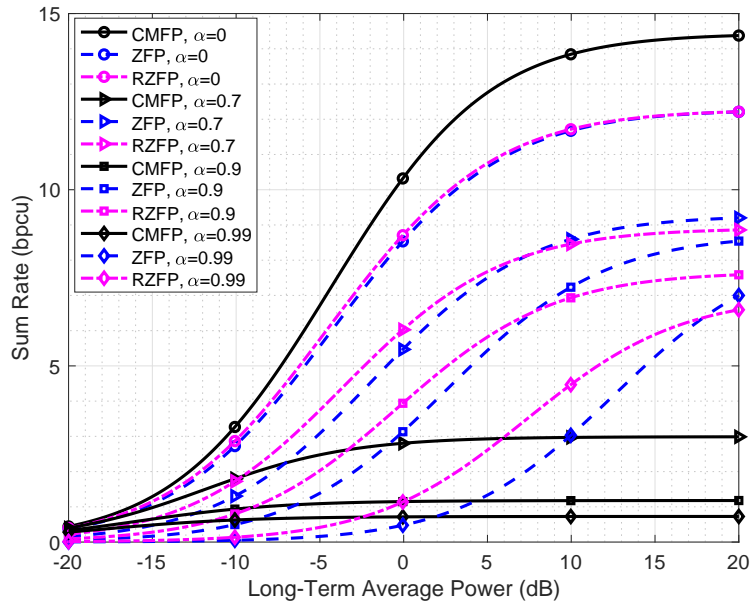}
\else
\includegraphics[width=6.5cm]{Figures/dl_2D_exp}
\fi
\vspace{-3.5mm}
\caption{Achievable sum rates of the three precoders in a downlink channel under the exponential correlation pattern with correlation parameter $\alpha \in \{0, 0.7, 0.9, 0.99\}$. The number of antennas at the base station is $M=8\times 8$, the number of users is $K=10$, the number of taps used in the computations is $L=4$, and the block length in the time domain is $T=100$. The distance between adjacent antenna elements is half the wavelength.}
\label{fig:dl_exp_2}
\end{figure}
Fig.~\ref{fig:dl_exp_2} shows the sum rate  performance in an exponential channel with parameter $\alpha \in \{0, 0.7, 0.9, 0.99\}$. The number of antenna elements is the same as in the case of ULA, $M=64$. The antenna elements are assumed to be in an $8\times 8$ configuration. Other parameters of the system are given in the figure caption. A comparison with Fig.~\ref{fig:dl_exp} shows that the performance for $\alpha=0$ is about the same. On the other hand, as expected, the performance degrades with $\alpha$. For example, at a long-term average power of 20 dB, with ULA, the sum rate with CMFP is approximately 6, 2.5, and 1 bpcu at $\alpha = 0.7, 0.9,$ and $0.99$ (Fig.~\ref{fig:dl_exp}), whereas with UPA, this sum rate drops to approximately 3, 1, and 0,8 bpcu for the same values of $\alpha$ (Fig.~\ref{fig:dl_exp}). Fig.~\ref{fig:dl_exp_2} shows that RZFP beats CMFP at a long-term average power value of 20 dB by 6, 6.5, and 5.5 bpcu at $\alpha = 0.7, 0.9,$ and $0.99.$ These values are at 3.5, 6, and 6 bpcu for ULA in Fig.~\ref{fig:dl_exp}.

\subsubsection{Downlink Channel: Bessel Correlation Pattern}
\begin{figure}[!t]
\centering
\ifCLASSOPTIONonecolumn
\includegraphics[width=11.5cm]{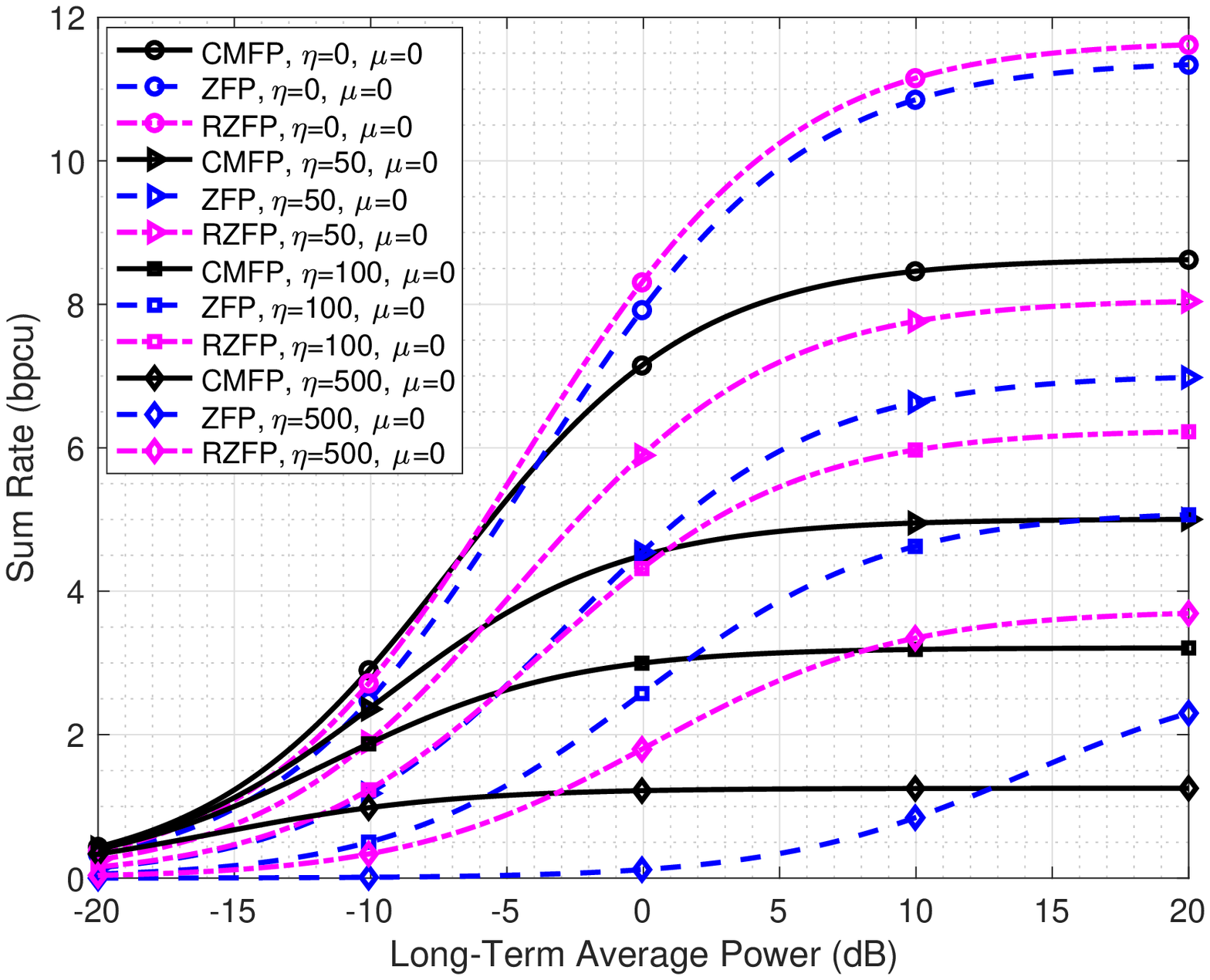}
\else
\includegraphics[width=6.5cm]{Figures/dl_2D_bsl}
\fi
\vspace{-3.5mm}
\caption{Achievable sum rates of the three precoders in a downlink channel under the Bessel correlation pattern with correlation parameters $\eta \in \{0, 50, 100, 500\}$ and $\mu = 0$. System parameters are $M=8\times 8$, $K=10$, $L=4$, $T=100$. The distance between adjacent antenna elements is half the wavelength.}
\label{fig:dl_bsl_2}
\centering
\ifCLASSOPTIONonecolumn
\includegraphics[width=11.5cm]{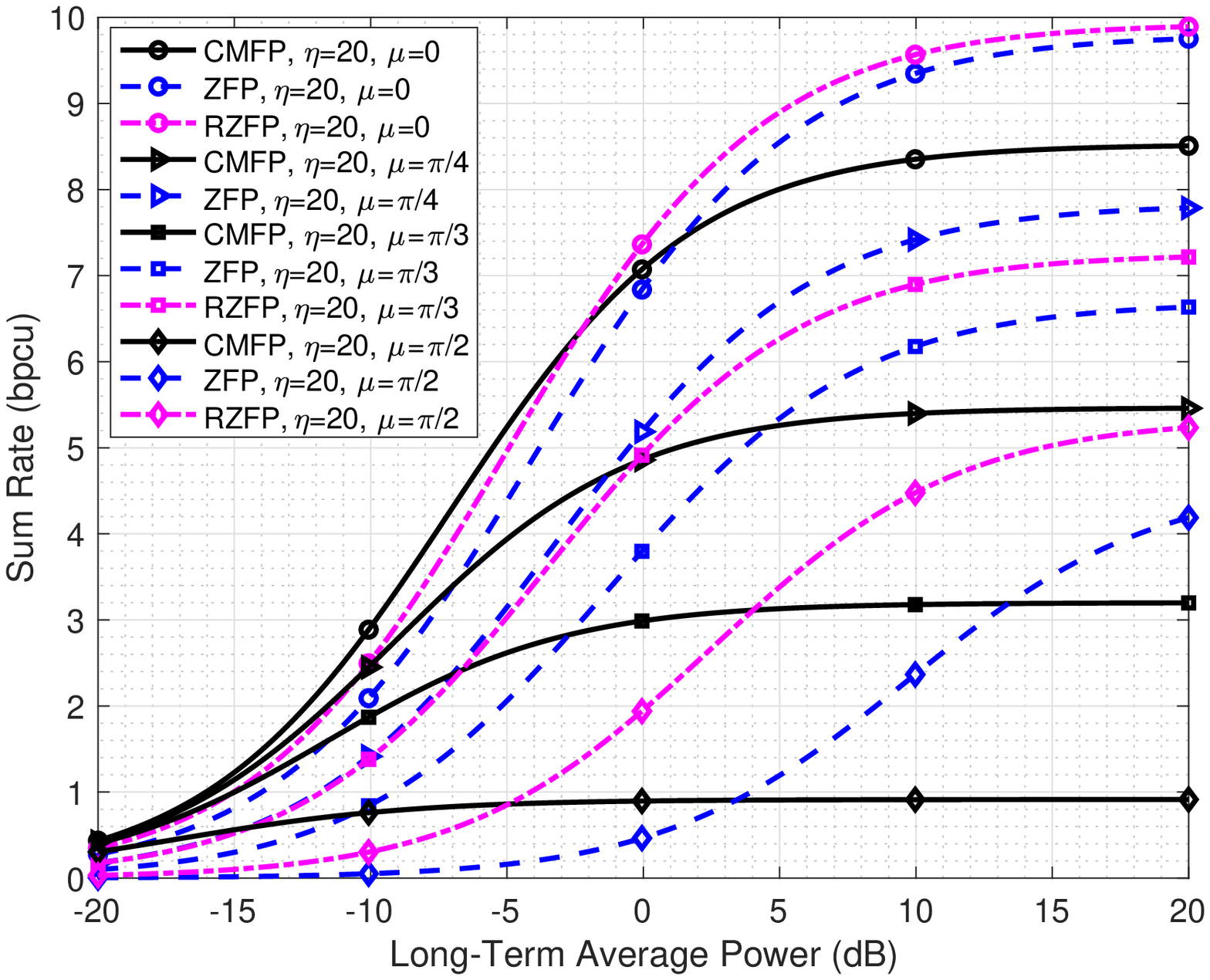}
\else
\includegraphics[width=6.5cm]{Figures/dl_2D_mu}
\fi
\vspace{-3.5mm}
\caption{Achievable sum rates of the three precoders in a downlink channel under the Bessel correlation pattern with correlation parameters $\mu \in \{0, \pi/4, \pi/3, \pi/2\}$ and $\eta = 20$. System parameters are $M=8\times 8$, $K=10$, $L=4$, $T=100$. The distance between adjacent antenna elements is half the wavelength.}
\label{fig:dl_mu_2}
\end{figure}
It was observed in Fig.~\ref{fig:dl_bsl} and Fig.~\ref{fig:dl_mu} that, in the case of ULA, the sum rate performance on the downlink for the Bessel correlation pattern was such that RZFP and ZFP outperformed CMFP for large values of average power. We observe that same phenomenon, but in a more significant fashion in Fig.~\ref{fig:dl_bsl_2} and Fig.~\ref{fig:dl_mu_2}. For example, in Fig.~\ref{fig:dl_bsl_2}, RZFP has about 2.5, 3, 3, and 2.5 bpcu better sum rate performance for $\eta=0,50,100, 500$ and $\mu=0$, respectively at a power level of 20 dB. These values were much smaller, around 1 bpcu or less, in the case of ULA in Fig.~\ref{fig:dl_bsl}. In a similar fashion, we observe from Fig.~\ref{fig:dl_mu_2} that RZFP is better than CMFP by about 1.5, 2, 4, and 4 dB at $\mu = 0, \pi/4, \pi/3$ and $\pi/2$ and $\eta = 20$ at a power level of 20 dB. Again, these values were smaller in the case of ULA. Note that, on the other hand, the sum rate values achieved after precoding are less in the case of UPA versus ULA. In both cases, ZFP performance is better than CMFP but not as good as RZFP. Fig~\ref{fig:dl_exp_2}--\ref{fig:dl_mu_2} verify our original expectations that the UPA performance will be lower than ULA due to higher correlation, but the amount with which RZFP and ZFP beat CMFP in the case of UPA will be higher than that of ULA.
\subsubsection{Uplink Channel: Exponential Correlation Pattern}
\begin{figure}
\centering
\ifCLASSOPTIONonecolumn
\includegraphics[width=11.5cm]{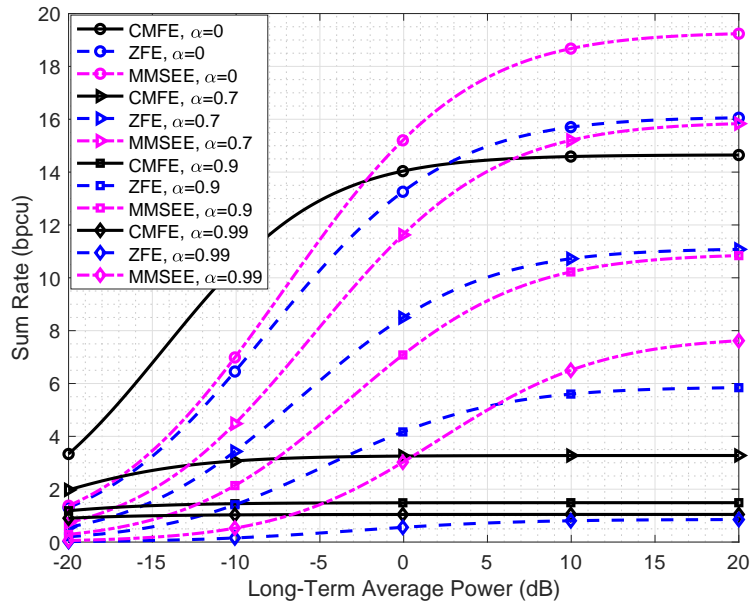}
\else
\includegraphics[width=6.5cm]{Figures/ul_2D_exp}
\fi
\vspace{-3.5mm}
\caption{Achievable sum rates of the three equalizers in the uplink channel under the exponential correlation pattern with correlation parameter $\alpha \in \{0, 0.7, 0.9, 0.99\}$. System parameters are $M=8\times 8$, $K=10$, $L=4$, $T=100$, $N=20$, $T_c=20$. The distance between adjacent antenna elements is half the wavelength.}
\label{fig:ul_exp_2}
\end{figure}

Fig.~\ref{fig:ul_exp_2} shows the sum rate performance of UPA with the exponential channel model on the uplink for $\alpha \in \{0, 0.7, 0.9, 0.99\}$. A comparison with Fig.~\ref{fig:ul_exp} shows the performance is about the same for $\alpha=0$. On the other hand, for $\alpha = 0.7, 0.9,$ and $0.99$, the behavior is similar to what was observed in Fig.~\ref{fig:ul_exp}. Namely, the sum rate performance degrades, but the gain by MMSEE over CMFE at sufficiently high power increases. We also note that, as in Fig.~\ref{fig:ul_exp}, the performance on the uplink for MMSEE is better than CMFE even for $\alpha=0$.

\subsubsection{Uplink Channel: Bessel Correlation Pattern}
\begin{figure}[!t]
\centering
\ifCLASSOPTIONonecolumn
\includegraphics[width=11.5cm]{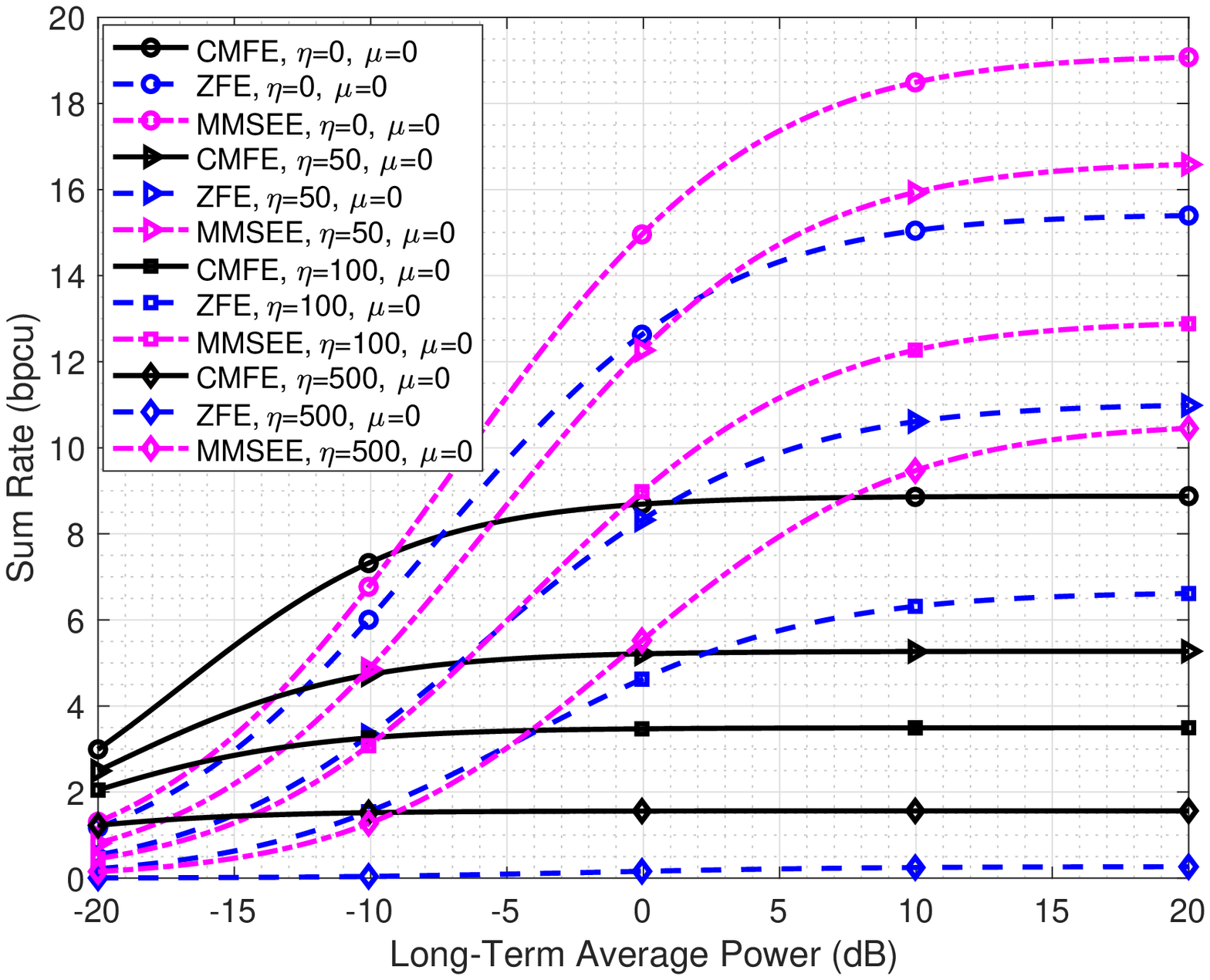}
\else
\includegraphics[width=6.5cm]{Figures/ul_2D_bsl}
\fi
\vspace{-3.5mm}
\caption{Achievable sum rates of the three equalizers in the uplink channel under the Bessel correlation pattern with correlation parameters $\eta \in \{0, 50, 100, 500\}$ and $\mu = 0$. System parameters are $M=8\times 8$, $K=10$, $L=4$, $T=100$, $N=20$, $T_c=20$. The distance between adjacent antenna elements is half the wavelength.}
\label{fig:ul_bsl_2}
\centering
\ifCLASSOPTIONonecolumn
\includegraphics[width=11.5cm]{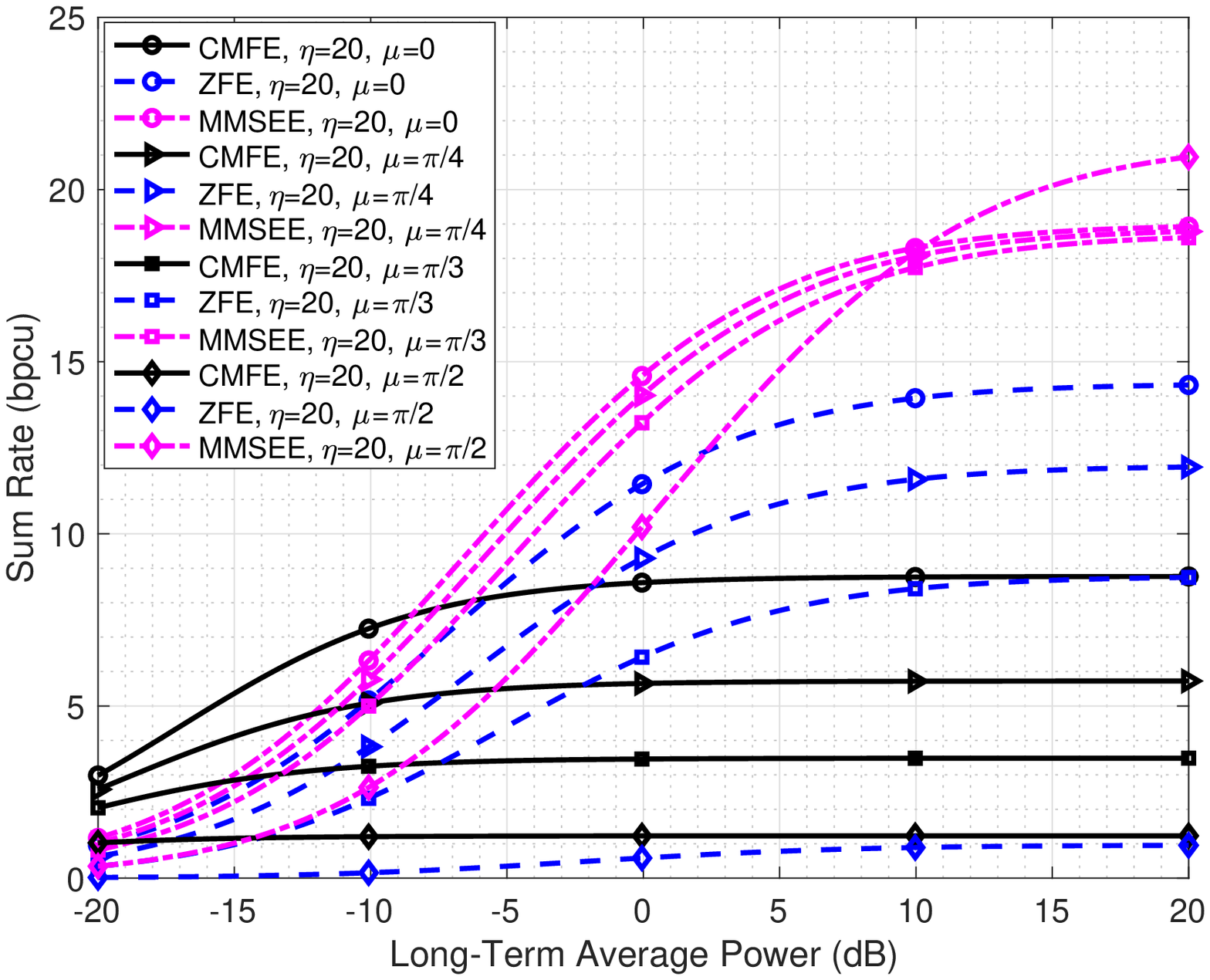}
\else
\includegraphics[width=6.5cm]{Figures/ul_2D_mu}
\fi
\vspace{-3.5mm}
\caption{Achievable sum rates of the three equalizers in the uplink channel under the Bessel correlation pattern with correlation parameters $\mu \in \{0, \pi/4, \pi/3, \pi/2\}$ and $\eta = 20$. System parameters are $M=8\times 8$, $K=10$, $L=4$, $T=100$, $N=20$, $T_c=20$. The distance between adjacent antenna elements is half the wavelength.}
\label{fig:ul_mu_2}
\end{figure}
We observe that the behavior in Fig.~\ref{fig:ul_bsl}--\ref{fig:ul_mu} is repeated in Fig.~\ref{fig:ul_bsl_2}--\ref{fig:ul_mu_2} in that the uplink channel has much better performance for MMSEE against CFME at the full long-term average power of 20 dB. In addition, with UPA, there is more performance degradation in the case of CMFE, while the gains by MMSEE against CMFE are higher. As stated earlier, the reason for this is the elimination of the IF term in the uplink at the BS, as discussed earlier in Section~\ref{sec:importanceofcsi}. In Fig.~\ref{fig:ul_bsl_2}, it can be observed that the performance of MMSEE with UPA is almost at the same level as the MMSEE performance with ULA in Fig.~\ref{fig:ul_bsl}.

\section{Conclusion}

As a result of the analysis and computations in this work, one can say that the sum rate of a channel is highly dependent on the nature of the channel. For uncorrelated  downlink channels, CMFP has optimal performance and it does perform near the upper bound of the channel. However, the presence of correlation among the antennas at the base station leads to a decrease in the achievable sum rate for the users in the system. Depending on the correlation model and the channel SNR, it is possible to employ another precoder to improve the sum rate performance for the downlink channel. In particular, RZFP and ZFP can outperform CMFP as far as the downlink channel is concerned.

Looking at the uplink results, one concludes that using MMSEE (which is equivalent of RZFP) and RZE (equivalent to ZFP) can improve the data rate in the channel. In addition, MMSEE is less sensitive to the correlation patterns considered in this paper.

We would like to state in concluding that long-term average power is small. Therefore, a mechanism to switch the precoder or the equalizer to CMFP or CMFE, respectively, will be a judicious action.

Two general observations are valid. First, the uplink channel is different than the downlink channel due to the presence of CSI. The BS takes advantage of this fact and performs well in the uplink, significantly better than the downlink, even in the face of correlation. Second has to do with the sum rate performance in the presence of UPA as compared to ULA. For UPA, due to a higher level of spatial correlations in the channel, sum rate performance degrades, but the gain with which RZFP and ZFP outperform CMFP as well as MMSEE and ZFE outperform CMFE increases, as can be expected.


\appendix

Let ${\bf F}_{(l,l-b)}[k,q]\triangleq d_l^{1/2}[k]{\bf h}_{l}[k]{\bf A}{\bf h}_{l-b}[q]d_{l-b}^{1/2}[q]$ and
$T \triangleq T_{l,l',b}[k,q] \triangleq\mathbb{E}\{ ({\bf F}_{(l,l-b)}[k,q])({\bf F}_{(l',l'-b)}[k,q])^*\}$.
Then,
\begin{align}
k=q, l=l', b=0 & \,\Rightarrow\, T =( {\rm tr}({\bf A}^2)+{\rm tr}^2 ({\bf A}) )d_l^2[k],\label{eqn:gmg1}\\
k=q, l=l', b\neq 0 & \,\Rightarrow\, T = {\rm tr}({\bf A}^2) d_l[k]d_{l-b}[k],\\
k=q, l\neq l', b = 0 & \,\Rightarrow\, T = {\rm tr}^2({\bf A})d_l[k]d_{l'}[k],\\
k=q, l\neq l', b\neq 0 & \,\Rightarrow\, T = 0\label{eqn:gmg4},\\
k\neq q, l=l', b=0 & \,\Rightarrow\, T= {\rm tr}( {\bf A}^2) d_l[k] d_l[q],\\
k\neq q, l=l', b\neq 0 & \, \Rightarrow \, T = {\rm tr} ({\bf A}^2) d_l[k] d_{l-b}[q],\\
k \neq q, l\neq l' & \, \Rightarrow \, T = 0.\label{eqn:gmg7}
\end{align}

\bibliographystyle{IEEEtran}
\bibliography{References/References}

\end{document}